A general parameterized mathematical food web model that predicts a stable green world in the terrestrial ecosystem

KOTARO KONNO

National Institute of Agrobiological Sciences, Tsukuba, Ibaraki 305-8634 JAPAN

Corresponding Author

Kotaro KONNO

Insect-Plant Interaction Research Unit

Division of Insect Sciences

National Institute of Agrobiological Sciences

1-2 Ohwashi, Tsukuba, Ibaraki 305-8634, Japan

Phone:+81-29-838-6087 Fax:+81-29-828-6028

E-mail:konno@affrc.go.jp

Abstract

Terrestrial ecosystems are generally green with vegetation and only a small part (<10%) of the total plant matter is consumed by herbivores annually, which means herbivore density is consistently low and stable. Nonetheless, the mechanism underlying this pattern has been unclear due to the lack of suitable food-web models for predicting the absolute density of the herbivore biomass in physical units. Here, I present a simple parameterized mathematical food-web model describing three trophic level systems that can predict the biomass density of herbivores **h** (kg protein/m$^3$) and carnivores **c** from ecological factors such as the nutritive values of plants $n_p$ (kg protein/m$^3$), herbivores $n_h$, and carnivores $n_c$, searching efficiency (volume) of carnivores S (m$^3$/m$^3$day=/day), eating efficiency (speed) of herbivores $e_h$ (m$^3$/m$^3$day=/day) and carnivores $e_c$, respiratory decrease in herbivore and carnivore biomasses, $d_h$ (kg/kg day=/day) and $d_c$, absorption efficiency of herbivores and carnivores $\alpha_h$ (ratio) and $\alpha_c$, and probabilities of carnivores preying on herbivores or carnivores, $P_{hc}$ (ratio) and $P_{cc}$. The model predicts a stable equilibrium with low herbivore biomass **h** sufficient to keep the world green provided the food-web consists of the three trophic levels, plants, herbivores and carnivores; intraguild predation of carnivores exists; $n_p<n_h,n_c$; S>>$e_h$; and $P_{hc}>P_{cc}>0$. These conditions are well-realized in above-ground terrestrial ecosystems where plant-rich "green world" is common, versus animal-rich belowground and aquatic ecosystems where some conditions are not realized. The **h** and **c** calculated from our model showed surprisingly good agreement with those from empirical observations in forest ecosystems, where both **h** and **c** are order of magnitude of ca. 100 mg (fresh biomass/m$^2$ forest), and in savannah ecosystems. The model predicts that the nutritive values of plants $n_p$, digestibility of plants by herbivores $\alpha_h$, and herbivore eating speed $e_h$ are positively correlated with **h** and the intensity of herbivory, which theoretically explains the out-door defensive effects of the anti-nutritive or quantitative defenses (e.g., tannins, protease inhibitors) of plants, and predicts that **c** and the carnivore/herbivore ratio **c/h** are positively correlated with the relative growth rate of herbivores $G_h$. The present model introduced parameterized realities into food-web theory, which were missing in previous models.

Key words: mathematical food web model; prediction of the absolute herbivore and carnivore biomass; parameterization; green world hypothesis; HSS hypothesis; nutritive value; searching efficiency; eating speed; relative growth rate; forest; savannah; plant defense

INTRODUCTION

Despite the existence of innumerable species of herbivores, most terrestrial regions of the world with adequate rainfall and temperature are covered with vegetation (i.e., green) and only a small part (ca. 1.5-11% in forests) of the plant materials or plant production are consumed by herbivores annually even though plant materials are abundant (Coley and Barone 1996; Cebrian 1999). This situation is sometimes called the "green world", and a number of hypotheses have been made to explain its causes (Polis 1999). One well-known explanation, the HSS hypothesis, posits that the consumption of herbivores by predators limits the number of herbivores and thereby their consumption of plants. Thus, while the population of plants, predators and decomposers are all resource-limited (i.e., limited by finite amounts of space, light, water, nitrogen, phosphorus and minerals), herbivores alone are subject to control by predators (Hairston et al. 1960). Considering the contrast between, for example, the fallen leaves on a forest floor, with their clearly low rates of consumption by herbivores, and the thousands of eggs laid by an individual moth in hopes that a few will escape consumption by a predator and reach maturity, the HSS hypothesis seems to offer a very reasonable explanation. However, the HSS only describes the current status, and does not explain why herbivores are controlled by predators (natural enemies) but plants are not controlled by herbivores. In other words, there has been no explanation why there are just a small population of herbivores that can consume small amounts of plant materials and why there is such a large population of predators enough to keep the population of herbivores small. To address these questions concerning the biomass or population of each trophic level and its stability, mathematical modeling of the food web systems will be needed. The classic Lotka-Volterra model treated the prey-prayer relationship mathematically (Volterra 1926; Lotka 1932), but this model does not have a stable convergent equilibrium. Further, it does not express ecological factors in clear and practical units such as kg/m$^3$ or kg/day, nor does it predict the practical abundance of each trophic level with units such as kg/m$^3$. Then, in order to clarify the conditions of food web stability in the predator-prey system, Rosenzweig and MacArthur invented a graphical description method based on isoclines of predators and prey, and by using this method, they predicted that the density-dependent limitations of resources for both predators and prey stabilize the food web equilibrium, while lowering or satiation of the carnivore ability caused by increased handling time of predators destabilizes the predator-prey system (Rosenzweig and MacArthur 1963). In keeping with this theory, Rosenzweig later predicted using a similar model that an enrichment of the ecosystem or an increase in nutrients would destabilize the predator-prey system (Rozenzweig 1971). They then further extended the two trophic-level model (predator-prey) into a three trophic-level model (carnivore-herbivore-plant), and predicted based on this three trophic-level model that intraspecific competition among plants and intraspecific competition among herbivores would stabilize the food web system (Rosenzweig 1973), whereas Wollkind predicted based on this model that negative interspecific and intraspecific carnivore interaction would stabilize the three trophic-level food web system (Wollkind 1976). Oksanen and his colleagues added the productivity of plants to this model and concluded that if the productivity of plants is large enough to compensate for the consumption of herbivores, the population of herbivores will become large enough to maintain predators, which will then control the herbivores; in this way a three trophic-level food web system emerges wherein carnivores control the population of herbivores to a relatively low level, making the effects of herbivory relatively small compared to the existent plant biomass, and maintaining a green ecosystem as in rain forests in tropical and temperate ecosystems with adequate temperature (Oksanen et al. 1981; Oksanen and Oksanen 2000). They also predicted that, when the productivity of plant is low, then the herbivore biomass or population will become too low to support the existence of carnivores. In this case a less green, two trophic-level system composed of only plants and herbivores emerges, wherein the population or biomass of plants is severely controlled by herbivores, as in tundra ecosystems in the far north (Oksanen 1981; Oksanen and Oksanen 2000). Their predictions are supported by empirical observations in mountain and tundra ecosystems (Okanen 1981; Oksanen and Oksanen 2000; Oksanen and Olofsson 2009). The above mathematical studies and models starting from Lotka-Volterra's model have elucidated various qualitative traits and characteristics of the food web structures, dynamics and stability. However, all these models, *per se*, provide a qualitative examination rather than a quantitative one linked to absolute values of ecological parameters, and do not predict the absolute abundance of the three trophic levels or the intensity of herbivory. The prediction of the absolute intensity of herbivory accompanied by physical units is indispensable to predict the color of the ecosystem, because if a stable equilibrium exists in a three trophic-level food web system and the intensity of herbivory is low enough for plants to tolerate or lower than plant productivity, then a stable green world system is expected to appear. In contrast, if the predicted herbivory intensity is too high for plants to tolerate or higher that plant productivity, then a stable equilibrium will not arise in the three trophic-level system, and the population of plants will be decreased by overherbivory, followed by a decrease in the herbivores themselves, and finally a collapse of the carnivore population due to the dearth of herbivores to prey on. These conditions may necessitate the transition to a two trophic-level system with only plants and herbivores, which is typical of tundra ecosystems as suggested by Oksanen et al. (Okanen 1981; Oksanen and Oksanen 2000; Oksanen and Olofsson 2009). The prediction of the absolute biomass of herbivores and carnivores, and the intensity of herbivory accompanied by physical units, are therefore important to predict the marginal conditions giving rise to a transition from a three trophic-level food web system, or "green world" system, to a two trophic-level food web system, or less "green world" system.

In most terrestrial ecosystems, the total animal biomass is small, an estimated ca. 30 - 640 (mg fresh mass/m$^2$) (Schowalter et al. 1981; Meyer et al. 2005), and reaches 8-20 (g fresh mass/m$^2$) only in extreme cases such as on the savannahs of Africa (e.g., the Serengeti) (Coe et al. 1976; Packer et al. 2005). The intensiveness of herbivory in terrestrial ecosystems, especially forests, is generally low (1-20%) (Coley and Barone 1996; Cebrian 1999), and terrestrial ecosystems are green (plant-rich), and never animal-rich. For example, in temperate forests in North America, the fresh biomass of chewing insects is expected to be somewhere around 120-360 (mg/m$^2$), while the fresh leaf biomass in forests is more than 300 (g/m$^2$) (Schowalter et al. 1981), and the biomass of herbivorous insects in a field composed of goldenrod, *Solidago gigantea*, in the wild in North America was estimated to be 6.24-50.4 (mg dry mass/m$^2$) or 31.2 -252 (mg fresh mass/m$^2$) (Meyer et al. 2005). In order to elucidate the reasons why the biomass of animals including herbivores is so small in terrestrial ecosystems, i.e., on an order of magnitude of 30-640 (mg fresh mass/m$^2$), it seems clear that a parameterized mathematical model or a theory that predicts the absolute amount of biomass of both herbivores and carnivores in realistic physical units such as (mg fresh biomass/m$^2$ or m$^3$ ecosystem) will be crucial. Nonetheless, few attempts have been made to establish such a model or theory; most of the existing models neither predict values with physical units nor include factors with physical units.

In this study, I present a simple parameterized mathematical food web model with a stable equilibrium that can predict the absolute abundance of biomass of each trophic level using realistic physical units. The model predicts the factors, mechanisms and conditions that realize a green (plant-rich) terrestrial world.

THE MATHEMATICAL MODEL AND THE EQUILIBRIUM

*The model setting of the food web structure and definitions of ecological factors*

The structure of the food web model is shown in Fig. 1, where all the variables are defined as in Table 1. The food web consists of three trophic levels, a plant (or plant material), an herbivore (or primary consumer), and a carnivore (or secondary consumer) level. Even though there are a large number of species in each trophic level, we regard each trophic level as a set. The total biomass densities (abundance) of herbivores (a set of herbivores) and carnivores (a set of carnivores) are expressed as $h$ (kg/m$^3$) and $c$ (kg/m$^3$). For biomass, we adopted dry mass of protein in this study because protein is the limiting factor for the growth of animals in many cases, but in other particular cases in which factors other than protein are stoichiometrically limiting, such as dry mass of phosphorus, nitrogen, iron, etc., these stoichiometrically limiting factors should be adopted as indicators of biomass. Since $h$ and $c$ are functions of time, they can be expressed as $h(t)$ and $c(t)$, respectively. Similarly, the ratios between the total volume (m$^3$) of herbivores and carnivores to the volume (m$^3$) of the particular ecosystem in which the herbivores and/or carnivores exist are expressed as $V_h(t)$ and $V_c(t)$, respectively. Plant (plant material), herbivore (primary consumer), and carnivore (secondary consumer) have their own nutritive values, $n_p$, $n_h$, and $n_c$ (kg/m$^3$ or kg protein/m$^3$).

I tentatively assume that the amount of available plants is ample for herbivores and that herbivores can eat as much as they wish (in a later section, I will examine whether this assumption is relevant after estimating the equilibrium solutions of herbivore and carnivore biomass **h** and **c** solving all the following equations), and therefore, $F_{out\,p}(t)$, the flow of biomass that goes out of the plant set per unit volume (m$^3$) per day, which is the plant material (biomass as protein) consumed by herbivores per unit volume per day, is expressed as

$$F_{out\ p}(t) = e_h n_p V_h(t) = e_h n_p \frac{h(t)}{n_h} \ (kg/m^3 day).$$

Here, $e_h$ (/day) is a feeding efficiency (or eating speed) defined as follows.

$$e_h := \frac{\text{volume of plant material that herbivores can eat per day}}{\text{volume of herbivores}} \ (/day)$$

When the absorption efficiency of herbivore feeding on plants is defined as follows,

$$\alpha_h := \frac{\text{plant biomass (as protein) absorbed by and incorporated into herbivores}}{\text{plant biomass (as protein) eaten by herbivores}} \ (\text{ratio and without unit}),$$

then, $F_{in\ h}(t)$, the flow of biomass (as protein) that is incorporated into the herbivore set per unit volume ecosystem per day resulted from herbivory is expressed as follows.

$$F_{in\ h}(t) = \alpha_h e_h n_p \frac{h(t)}{n_h} \ (kg/m^3 day)$$

$D_h(t)$, the amount of biomass (as protein) consumed by herbivores in the form of respiration, metabolism, and excretion per unit volume ecosystem is expressed using the decreasing constant $d_h$ as follows.

$$D_h(t) = d_h h(t) \ (kg/m^3 day)$$

If carbohydrates are selectively consumed by herbivores for respiration, then $d_h$ will be much smaller (because the model focuses on the flow of protein).

I then assume that the preying behavior of carnivores is as follows. Carnivores with volume V can search for volume SV in unit time (day) for prey, and if prey herbivores exist in this volume, then the carnivores will eat herbivores within this area at the preying probability of $P_{hc}$ (ratio and without unit). Here, the constant S (/day) is the searching efficiency of carnivores. Then, $F_{out\,h}(t)$, the flow of biomass that goes out of the herbivore set per unit volume ecosystem per day, which is the biomass (as protein) of herbivores consumed by carnivores per unit volume ecosystem per day is expressed as follows.

$$F_{out\,h}(t) = P_{hc} n_h V_h(t) S V_c(t) = \frac{P_{hc} h(t) c(t) S}{n_c} \quad (kg/m^3 day)$$

$F_{in\,c}(t)$, the biomass (as protein) of herbivores incorporated into carnivores per unit volume ecosystem per day as a result of carnivory on herbivores is expressed as follows using the absorption efficiency of carnivores feeding on animal biomass (herbivores and carnivores) $\alpha_c$ defined in a similar way as $\alpha_h$.

$$F_{in\,c}(t) = \frac{\alpha_c P_{hc} h(t) c(t) S}{n_c} \quad (kg/m^3 day)$$

$D_c(t)$, the amount of biomass (as protein) consumed by carnivores in the form of respiration, metabolism, and excretion is expressed using the decreasing constant $d_c$ as follows.

$D_c(t) = d_c c(t)$ (kg/m³day)

The present model has uniqueness in that the model assumes intraguild predation of carnivores, the existence of which is manifest in most real ecosystems but has often been neglected in many food web models, including the Lotka-Voltera model. In some models, intra-carnivore interactions, negative or positive, are included, such as in the model of Wollkind (Wollkind 1976), but in the present model, intraguild predation of carnivores is explicitly defined as follows. Carnivores with a local volume density (ratio) of $V_c(t)$ will search the area (ratio) of $SV_c(t)$ per unit time (day) and consume the carnivores that exist in that area (ratio) at the preying probability of $P_{cc}$. Then, $F_{out\ cc}(t)$, the biomass of carnivores consumed by other carnivores per unit volume ecosystem (m³) per unit time (day) can be expressed as follows.

$$F_{out\ cc}(t) = \frac{P_{cc} n_c V_c(t) S V_c(t)}{2} = \frac{P_{cc} S c(t)^2}{2 n_c} \text{ (kg/m}^3\text{day)}$$

The reason for the factor 1/2 is to avoid the double counting of intraguild predation events (i.e., carnivore A cannot prey on carnivore B while carnivore B is preying on carnivore A). $F_{in\ cc}(t)$, the biomass of carnivores predated by other carnivores and incorporated into these carnivores is expressed as follows using absorption efficiency $\alpha_c$.

$$F_{in\ cc}(t) = \frac{\alpha_c P_{cc} S c(t)^2}{2 n_c} \text{ (kg/m}^3\text{day)}$$

Therefore, $F_{out\ c}(t)$, the overall loss of biomass from the set of carnivores through the events of intraguild predation of carnivores is written as follows.

$$F_{out\ c}(t) = F_{out\ cc}(t) - F_{in\ cc}(t) = \frac{P_{cc}Sc(t)^2}{2n_c} - \frac{\alpha_c P_{cc}Sc(t)^2}{2n_c} = \frac{(1-\alpha_c)P_{cc}Sc(t)^2}{2n_c} \ (kg/m^3 day)$$

*The equilibriums of biomass concentrations of herbivores and carnivores*

The differential equations describing the biomass density of herbivores $h(t)$ and the biomass density of carnivores $c(t)$ derived from the above model are described as follows.

$$\frac{dh(t)}{dt} = F_{in\ h}(t) - D_h(t) - F_{out\ h}(t) = \frac{\alpha_h e_h n_p h(t)}{n_h} - d_h h(t) - \frac{P_{hc}Sh(t)c(t)}{n_c} \quad (1)$$

$$\frac{dc(t)}{dt} = F_{in\ c}(t) - D_c(t) - F_{out\ c}(t) = \frac{\alpha_c P_{hc}Sh(t)c(t)}{n_c} - d_c c(t) - \frac{(1-\alpha_c)P_{cc}Sc(t)^2}{2n_c} \quad (2)$$

Under the equilibrium condition where both $\frac{dc(t)}{dt}$ and $\frac{dc(t)}{dt}$ are 0, the equilibrium solutions of herbivore biomass **h** and carnivore biomass **c** are obtained from equations (1) and (2) as follows.

$$\mathbf{h} = \frac{n_c\{(1-\alpha_c)P_{cc}(\alpha_h e_h n_p - n_h d_h) + 2n_h d_c P_{hc}\}}{2\alpha_c P_{hc}^2 n_h S} \ (kg/m^3) \quad (3)$$

$$\mathbf{c} = \frac{n_c(\alpha_h e_h n_p - n_h d_h)}{P_{hc} n_h S} \ (kg/m^3) \quad (4)$$

These two formulas, (3) and (4), are the major results of the present study, and have many implications for the food web structures of a variety of ecosystems.

Using $G_h$, the relative (internal) growth rate of the herbivore set continuously eating plant material that can be defined as

$$G_h =: \frac{\alpha_h e_h n_p - n_h d_h}{n_h},$$

equations (3) and (4) can be expressed as follows.

$$h = \frac{n_c\{(1-\alpha_c)P_{cc}G_h + 2d_c P_{hc}\}}{2\alpha_c P_{hc}^2 S} \quad (kg/m^3) \ (3)'$$

$$c = \frac{n_c G_h}{P_{hc} S} \quad (kg/m^3) \ (4)'$$

*The stability of the equilibrium and its conditions*

a. Convergence

Unlike in the Lotka-Voltera model, the above equilibrium (3), (4) in our model is generally stable and convergent, which can be easily demonstrated as follows. The differential equations (1), (2) can be simplified as

$$\frac{dh(t)}{dt} = Ah(t) - Bh(t)c(t) \qquad (1)'$$

$$\frac{dc(t)}{dt} = -Cc(t) - Dc(t)^2 + Eh(t)c(t) \qquad (2)'$$

with all A, B, C, D, E > 0 , when A, B, C, D, E are defined as

$$A := \frac{\alpha_h e_h n_p}{n_h} - d_h \qquad B := \frac{P_{hc} S}{n_c} \qquad C := d_c \qquad D := \frac{(1-\alpha_c)P_{cc} S}{2n_c} \qquad E := \frac{\alpha_c P_{hc} S}{n_c}.$$

It is clear by definition that B, C, D, and E are positive values, but it is also clear that A is a positive value because unless A is positive, herbivores cannot even grow under the ideal conditions in which they are permitted to eat plant material at will.

The linear approximation around the equilibrium

$$\mathbf{h} = \frac{AD+BC}{BE} \quad (3)"$$

$$\mathbf{c} = \frac{A}{B} \quad (4)"$$

is described as follows.

$$\begin{pmatrix} \frac{d\Delta h(t)}{dt} \\ \frac{d\Delta c(t)}{dt} \end{pmatrix} = \begin{pmatrix} 0 & -\frac{AD+BC}{E} \\ \frac{AE}{B} & -\frac{AD}{B} \end{pmatrix} \begin{pmatrix} \Delta h(t) \\ \Delta c(t) \end{pmatrix} \quad (5)$$

In order to examine whether the equilibrium is stable and convergent around the equilibrium point, the eigenvalue $\lambda$ of the above square matrix is achieved by solving the following equation.

$$\det \begin{pmatrix} -\lambda & -\frac{AD+DC}{E} \\ \frac{AE}{B} & -\frac{AD}{B} - \lambda \end{pmatrix} = \lambda^2 + \frac{AD}{B}\lambda + \frac{A(AD+BC)}{B} = 0 \quad (6)$$

Therefore, the solution for the eigenvalue $\lambda$ is as follows.

$$\lambda = \frac{-\frac{AD}{B} \pm \sqrt{\left(\frac{AD}{B}\right)^2 - \frac{4A(AD+BC)}{B}}}{2} \quad (7)$$

Since all A, B, C, D, and E are positive constants, it is clear from formula (7) that the real part of the solutions λ is always negative, whether or not λ is a real number or complex (imaginary) number (more likely under practical conditions), and this result means that the equilibria **h** and **c** obtained as formulas (3) and (4) are stable and convergent (likely to be rotatively convergent). The convergence of the present model is in clear contrast to the classic Lotka-Volterra model in which the equilibrium is neutrally stable but not convergent. This difference is caused mainly because our model includes intraguild predation of carnivores (i.e., $P_{cc} > 0$ is included in the present model), while the model of Lotka-Volterra does not. Therefore if $P_{cc}=0$ is assumed in the present model (as in the Lotka-Volterra model), then $D = 0$ is realized and the solution

$$\lambda = \pm\sqrt{ACE}\ i \quad (8)$$

is realized, while $i$ is an imaginary unit. This result means that if intraguild predation of carnivores is assumed to be absent ($P_{cc}=0$) in the present model, the equilibrium becomes neutrally stable but not convergent, and the densities of both carnivores and herbivores keep oscillating (rotating) around the equilibrium just as in the Lotka-Volterra model. Therefore it can be said that in ecosystems, intraguild predation of carnivores stabilizes the food web system and makes the equilibrium convergent, although an intraguild predation that is too large will increase the biomass of herbivore **h** and may finally cause an outbreak of herbivores, which will be examined later in this study. It is now clear from the above discussion that the equilibrium is generally stable and convergent in the present model except in the case described below.

b. Satiation of carnivores and outbreak of herbivores

The differential equations (1) and (2) are, in form, structurally similar to those presented in previous models (Rosenzweig and MacArthur 1963; Rosenzweig 1971; Oksanen and Oksanen 2000), but there are also some important differences: while the previous models include the effect of handling time in the differential equations, the present model does not. Instead, the present model assumes the factor $e_c$ (feeding efficiency or maximum eating speed), which is determined from factors including the capacity of digestive organs, the time needed for digestion absorption and assimilation, and the handling time (*sensu stricto*) of herbivores. From the physiologists' point of view, the rate limiting factors for $e_c$ of carnivores would be digestive and assimilation processes rather than handling processes (*sensu stricto*) under ordinary circumstances. Then by adopting $e_c$, the inequality conditions (10), (10)', (11), and (11)' can be set in order to judge whether carnivores are satiated as follows.

The equilibrium (3) and (4), however, is not realized when carnivores are satiated with food (i.e., herbivores and carnivores that carnivores encounter) and carnivores cannot function as expected to control the biomass of herbivores, leading to an outbreak of herbivores. The satiation of carnivores occurs under the following conditions, when the amount of available food (herbivores + carnivore) is larger than the eating ability (maximum speed) of the carnivores.

$$e_c \mathbf{V_c} < P_{hc} \mathbf{V_h} S \mathbf{V_c} + P_{cc} \mathbf{V_c^2} S \quad (9)$$

Here, $e_c$ (/day) is the feeding efficiency (or the maximum eating speed) defined as follows.

$$e_c := \frac{\text{volume of animal material that carnivores can eat per day}}{\text{volume of carnivores}} \quad (/\text{day})$$

and $\mathbf{V_h}$ and $\mathbf{V_c}$ are $V_h(t)$ and $V_c(t)$ under the equilibrium condition, respectively.
Then using the relationships

$$V_h = \frac{h}{n_h} \qquad V_c = \frac{c}{n_c}$$

and formulas (3) and (4), the inequality (9) indicating the condition of satiation is written as follows.

$$e_c < \frac{\{n_c(1-\alpha_c)+2\alpha_c n_h\}P_{cc}(\alpha_c e_h n_p - n_h d_h)}{2\alpha_c P_{hc} n_h^2} + \frac{n_c d_c}{\alpha_c n_h} \qquad (10)$$

or

$$e_c < \frac{\{n_c(1-\alpha_c)+2\alpha_c n_h\}P_{cc} G_h}{2\alpha_c P_{hc} n_h} + \frac{n_c d_c}{\alpha_c n_h} \qquad (10)'$$

When the condition (10) is satisfied, then the stable equilibrium described in formulas (3) and (4) will not be realized, the food web system will become less stable and an outbreak of herbivores will be more likely to happen.

On the other hand, when the condition

$$e_c > \frac{\{n_c(1-\alpha_c)+2\alpha_c n_h\}P_{cc}(\alpha_c e_h n_p - n_h d_h)}{2\alpha_c P_{hc} n_h^2} + \frac{n_c d_c}{\alpha_c n_h} \qquad (11)$$

or

$$e_c > \frac{\{n_c(1-\alpha_c)+2\alpha_c n_h\}P_{cc} G_h}{2\alpha_c P_{hc} n_h} + \frac{n_c d_c}{\alpha_c n_h} \qquad (11)'$$

is satisfied, then the ecosystem (or food web) will certainly remain in stable and convergent equilibrium as described in formulas (3) and (4).

PREDICTED VALUES FROM THE MODEL AND COMPARISON WITH EMPIRICAL DATA FROM SEVERAL TYPES OF ECOSYSTEMS: IS A "GREEN WORLD" SYSTEM REALIZED IN THE TERRESTRIAL ECOSYSTEM?

*General settings*

Next, I will examine whether the herbivore density in the stable equilibrium denoted as

$$h = \frac{n_c\{(1-\alpha_c)P_{cc}(\alpha_h e_h n_p - n_h d_h) + 2n_h d_c P_{hc}\}}{2\alpha_c P_{hc}^2 n_h S} \text{ (kg/m3)} \qquad (3)$$

is low enough to keep the terrestrial world green.

Here, I set the ecosystem using ecologically reasonable numerical values according to the previous reports (Schoonhoven et al. 2005; Begon et al. 2006). Leaves are distributed within the leaf layer with a depth of 1 m and a leaf area index (LAI) of 4, and the mass of leaves is 100 (g fresh mass/m² leaf). These conditions simulate grassland or forest canopy in which 1m³ of ecosystem contains 400g of fresh (wet) plant leaf material. Plant leaf materials contain 2% protein per fresh mass ($n_p=2\times10^1 \text{kg/m}^3$), and animal materials (herbivores and carnivores) contain 20% protein per fresh mass ($n_h=n_c=2\times10^2 \text{ kg/m}^3$). Since plant materials are more difficult to digest than animal materials, I assume that $\alpha_h=0.6$ and $\alpha_c=0.8$. The factors $P_{hc}$ (the probability of herbivores being predated by carnivores once the herbivores have entered the searching volume (area) of carnivores), $P_{cc}$ (the probability of carnivores being predated by other carnivores after entering one another's searching volume (area)), and S (the relative searching area per unit time), the three factors related to predation behavior, are very difficult to estimate, and in fact there have been very few empirical observations of these factors. The assessment of these factors is further complicated by the fact that some carnivores (bugs, ants, etc.) search for prey over the surface, which is 2-dimensional, while other carnivores (flying insects, birds and planktons) search over space, which is 3-dimensional. While leaves are flat and could be treated as 2-dimensional, they exist in 3-dimensional space in the leaf canopy. Similarly, while some ecological studies treat ecosystems as two-dimensional areas and thus express biomass and other ecological factors using units of area (e.g., $mg/m^2$), others treat ecosystems 3-dimensionally and express biomass and other factors per unit volume (e.g., $mg/m^3$), which may have hampered quantitative studies in both practical and theoretical fields of ecology. Because the present model treats ecosystems 3-dimensionally, I have converted 2-dimensional measures into 3-dimensional ones based on the following principles. Since leaves scatter within a life zone of approximately 1 m depth in most ecosystems, including grasslands and forest canopies (i.e., the thickness of the canopy and grassland can be assumed to be approximately 1 m in most ecosystems), and the LAI is around 4 in many ecosystems, I converted 4 $m^2$ of leaf area to 1 $m^2$ of 2-dimensional area in the field, which could be further converted to 1 $m^3$ of volume in space for use in the present model. I therefore set $P_{hc}=0.25$ and $P_{cc}=0.05$.

*Herbivorous insects as herbivores*

a. Estimation from a Wasp - Caterpillar System in the forest

Here, I first assume that both herbivores and carnivores are ectotherms (cold-blooded animals such as insects and reptiles), and set the decrease of animal mass caused by respiration, metabolism and excretion as $d_h=d_c=0.006$ (/day). For eating efficiency $e_h=5$ (/day assuming caterpillars) and $e_c=1$ (/day assuming insect carnivores) are set. Then $G_h=0.296$ (/day) is obtained, which agrees well with the relative growth rates observed for caterpillar feeding on leaf material (Schoonhoven et al. 2005). In order to calculate **h** and **c** using our model, S is the most important value to estimate, although this estimation is difficult. According to field experiments on the preying ability of an insectivorous wasp, *Polites chinensis*, preying on *Pieris rapae* larvae in a cabbage field (Morimoto 1960), I estimated $S = 1.298 \times 10^7$ (/day), because the study showed that 28 worker wasps with a weight of 0.1g each searched a cabbage field of 65 m$^2$, and the number of *Pieris* butterfly larvae decreased from 93 to 80 (a 14% decrease) after 1 day of searching, while $P_{hc}=0.25$. Then, **h**, the biomass of herbivores per unit area m$^2$ or unit volume m$^3$ calculated from equation (3) could be determined as follows.

**h** $= 9.15 \times 10^{-7}$ (kg protein/m$^3$) $= 0.915$ (mg protein/m$^3$) $= 4.58$ (mg real fresh mass of caterpillars/ m$^2$) $= 22.9$ (mg apparent fresh mass with 80% of food material inside the gut of caterpillars weighed together/m$^2$; most ecological studies make this measurement with the contents of the guts measured together) (Table 2). Then the annual consumption of leaf volume is calculated as 4.58 (mm$^3$/m$^2$) × 5 ($e_h$) × 200 (days) = 4.58 (cm$^3$/m$^2$ year) =4.58 (g/m$^2$ year). Since we assume that 400 (g/m$^2$) of leaves exist, the estimated annual consumption rate of leaves by herbivores (caterpillars) is 1.14% (Fig. 2).

Similarly, the biomass density of carnivore **c** can be calculated from equation (4) as **c** = $1.812 \times 10^{-5}$ (kg protein/m$^2$) = 18.1 (mg protein /m$^2$) = 90.6 (mg fresh mass /m$^2$) (Fig. 2). The right side of the inequality is calculated to be 0.07365, and it is manifest that $e_c \gg 0.076$, and therefore the satiation of carnivores is not likely and the equilibrium is stable and convergent under this condition.

b. Estimation from a Carnivorous Bug – Herbivorous Beetle System in the forest

Here I estimated S from the empirical observation of a carnivorous bug, *Podius maculiventris* (Heteroptera), preying on the larvae of the Mexican bean beetle, *Epilachena varivestis* (Coleoptera) (Wiedenmann and O'Neil 1991). According to this studym bugs with a fresh mass of 60 (mg/individual) can search 0.20 m$^2$ of leaf surface, which can be translated to 0.05 m$^3$ of leaf canopy in 1.13 hr, and individual bugs are estimated to be searching for prey 5 hr/day. Then S can be calculated as S = 3.69 × 10$^6$ (/day). The biomass of herbivore **h** is then calculated as **h** = 3.22 × 10$^{-6}$ (kg protein/m$^3$(or m$^2$)) = 3.22 (mg protein/ m$^3$(or m$^2$)) = 16.1 (mg real fresh mass/m$^3$(m$^2$)) = 80.6 (mg apparent fresh mass including gut content/m$^3$ (m$^2$)) (Table 2). When we assume that herbivores eat 200 days/year and e$_h$=5, then it can be estimated that herbivores consume 16.1 (g/year/m$^3$(m$^2$)) in total, and in an ecosystem with a leaf biomass of 400 (g fresh mass/m$^2$), the annual consumption rate of leaves by herbivores is estimated to be 4.03%. The biomass of carnivore **c** is calculated as **c** = 6.39 × 10$^{-5}$ (kg protein/m$^3$(m$^2$)) = 63.7 (mg protein/m$^3$(m$^2$)) = 319 (mg fresh mass/ m$^3$(m$^2$)) (Table 2). Since the right side of the inequalities (10) and (11) is 0.07365, and it is apparent that e$_c$>>0.07356, the satiation of carnivores is unlikely and the equilibrium is stable and convergent.

c. Estimation from a Bird – Caterpillar System in the forest

In this system, the carnivores are birds, and because birds are endothermal, I assumed d$_c$ = 0.03, while d$_h$ and all other factors except S remained unchanged as in the Wasp - Caterpillar System and Carnivorous Bug – Herbivorous Beetle System above. Empirical observation of the hunting behavior of birds (Royama 1970) showed that two adult birds of the Great tit species, *Parus major*, carried 600 *Tortrix viridana* individuals (Tortricidae, Lepidoptera) to their shared nest every day from the oak forest (*Quercus petraea*) in Wytham Woods near Oxford in the UK, while the number of *Tortrix viridana* individuals per 1000 oak leaves (ca. 10 m² of leaves) was ca. 20. Since I assume that $P_{hc}$ = 0.25 and LAI = 4, and that the body volume of the great tit is 20 cm², S is calculated as S = 7.5 × 10⁶ (/day). Then **h** is calculated using equation (3) as **h** = 4.78 × 10⁻⁶ (kg protein/m³(m²)) = 4.78 (mg protein/m³(m²)) = 23.92 (mg real fresh mass caterpillar/m³(m²)) = 119.6 (mg apparent fresh mas caterpillar including gut content/m³(m²)) (Fig. 2). Then the annual consumption of leaf volume is calculated as 23.92 (mm³/m²) × 5 ($e_h$) × 200 (days) = 23.92 (cm³/m³(m²) year) =23.92 (g/m3(m²) year) (Table 2). Since we assumed the existence of 400 (g/m²) of leaves, the estimated annual consumption rate of leaves by herbivores (caterpillars) is 5.98%. Similarly, the biomass density of carnivore **c** can be calculated from equation (4) as **c** = 3.131 × 10⁻⁵ (kg protein/m²) = 31.3 (mg protein /m²) = 156.8 (mg fresh mass /m²) (Table 2). Since the right side of the inequalities (10) and (11) is 0.10365, and it is likely that $e_c$ > 0.10365, the satiation of carnivores is unlikely and the equilibrium is stable and convergent.

d. Comparison between the predicted values and empirical data of biomasses and annual leaf consumption in forest and grassland ecosystems where insects are the main herbivores

The present model predicted that the annual consumption rates of leaves by insect herbivores were 1.14%, 4.03%, and 5.98% in the Wasp-Caterpillar, Herbivorous Bug Caterpillar, and Bird-Caterpillar systems, respectively (Table 2). An intensive survey of numerous natural sites showed that the annual consumption rates compared to primary production to be 0-14% (quartile 25-75% to be 0-6%, average 1.5%) in forests and shrublands together, and 0-65% (quartile 25-75% to be 1-43%, average 23%) in grasslands (Cebrian 1999) (Table 2). The predictions of annual consumption rate made by the present model fit well within the range of annual consumption rates obtained from empirical observations in forests and shrublands. The present model predicts that the biomasses of herbivore **h** are 22.9, 80.6, and 119.6 (mg apparent fresh mass including gut content/$m^3$ ($m^2$)) for the Wasp-Caterpillar, Carnivorous Bug-Herbivorous Beetle, and Bird-Caterpillar systems, respectively (Table 2). Surprisingly, these predicted values were on the same order of magnitude as the empirical data—e.g., 120-360 (mg fresh biomass of chewing herbivores (including gut content)/$m^2$ forest) obtained in intensive surveys performed in North Carolina forests in the US (Schowalter et al. 1981) and 31.2-252 (mg fresh biomass of herbivorous insects (including gut content)/$m^2$) observed in stands of *Solidago gigantea* in North America (Meyer et al. 2005) (Table 2), although the predicted values tend to be somewhat smaller than the empirical ones. The biomass values for carnivore **c** predicted by the present model were 90.6, 319, and 156.8 (mg fresh mass/ $m^3$($m^2$)) in the Wasp-Caterpillar, Carnivorous Bug-Herbivorous Beetle, and Bird-Caterpillar systems. Here again, these predicted values were surprisingly on the same order of magnitude as the empirical data—e.g., 40 - 240 (mg fresh biomass of arthropod predators/$m^2$ forest) obtained in intensive surveys performed in North Carolina forests in the US (Schowalter et al. 1981), 100 - 376 mg fresh biomass of *Vespula* wasps introduced into a New Zealand beech forest (Thomas et al. 1990), and ca. 20-30 mg fresh biomass of insectivorous birds in tropical and temperate forests (Ternorgh et al. 1990; Holmes and Sherry 2001) (Table 2). However, it should be noted that the real biomass of carnivores may be somewhat larger due to the elusiveness of carnivores, and some groups of carnivores, including ants, wasps, reptiles, and amphibians, may not be counted in some measurements. Although the predicted values and empirical observation were on the same order of magnitude, it seems that the predicted values for biomass of herbivores **h** and annual consumption rates were several times smaller than their empirical counterparts. In the future, it will be necessary to examine the underlying cause of this discrepancy, but it is possible that this discrepancy was caused by the fact that the estimation of searching ability-related factors of carnivores, $P_{hc}$ and S is estimated from the combination of herbivores and carnivores that have sizes suitable for prey-predator interactions. In real ecosystems, some preys are too small or too large for predators to prey on. Therefore the estimation for $P_{hc}$ (and also $P_{cc}$) may be smaller than those estimated here. Predators may have the ability to prey on animals with individual biomasses with a range of say 2 orders of magnitude, and preys in a forest ecosystem have individual biomasses with a range of 5 orders of magnitude (0.1 mg to 10 g), and therefore the probability for the predation incidences $P_{hc}$ and $P_{cc}$ may be several-fold smaller, which could result in several-fold higher estimations for the **h**, **c**, and annual consumption rates.

*Herbivorous mammals as herbivores*

Estimation for the Lion-Wildebeest (or Zebra) system on the African savannah

For herbivorous mammals, the following reasonable ecological values are assumed according to the data and descriptions from various sources on ecology and stock raising. The apparent fresh mass of herbivorous mammal individuals is set to be 200 (kg fresh mass/individual wildebeest or zebra), which includes 60 kg of gut content (grasses), assuming that each herbivore individual takes 20 kg of fresh grass/day and the grass stays in the gut for three days for digestion. Therefore the real wet biomass for an herbivore individual is assumed to be 140 (kg real fresh mass/individual), and the estimated $e_h$ is 0.143. For carnivores (lions), I assume an individual biomass of 200 (kg fresh mass/individual). Since an average adult lion needs to consume 6 (kg/day) of meat to maintain its body mass, and since the protein composition of the meat (prey) is the same as that of a lion itself, it is reasonable to assume $d_c = 0.03$. For herbivorous mammals, however, the estimation of $d_h$ is more difficult. It would be reasonable to assume the same rate of energy consumption in both carnivorous mammals and herbivorous mammals, but herbivorous mammals consume a diet rich in carbohydrates derived from cellulose, and may utilize carbohydrates as a source of energy production rather than proteins, which they may save for growth of the body. Because our model is based on the flow and flow balance of proteins, $d_h$ would be expected to be much smaller than 0.03 as in carnivorous mammals ($d_c$), but it is still difficult to estimate $d_h$ and also ($\alpha_h e_h n_p - n_h d_h$) in equation (3) and (4). Therefore, I used $G_h$ and equations (3)' and (4)' to calculate **h** and **c**. The estimation of $G_h$, a relative (internal) growth rate of the herbivore set that freely eats grass, is made according to the realistic assumption that each female herbivorous mammal individual raises 1 progeny once in 2 years. This setting means that it takes 4 years (1460 days) for the population and biomass of herbivorous mammals to double under ideal conditions, taking the existence of male individuals into account. If $G_h$ is calculated from these ideal conditions, it would be $\sqrt[1460]{2}$ =0.0004748 (/day), but under real conditions in the field, $G_h$ for the whole herbivore set would be lower than this, because not all females are productive (some young and old females are not productive), and furthermore, the proportion of infants, which are still growing and are the sole source of increase of biomass in the herbivore set, would be smaller than expected from ideal conditions due to the selective predation of younger individual by carnivores. Taking these factors into account, $G_h$ = 0.0002 (/day) is assumed. The searching efficiency of herbivore S (/day) is estimated as follows. The first very important thing to know about this system is that herbivorous mammals such as wildebeests and zebras can run faster than carnivorous mammals such as lions (Elliott et al. 1977), which is clearly a very different situation compared to the above-described systems involving caterpillars. This means that the lion should depend on a sit-and-wait-type strategy, and that the hunting of carnivorous mammals waiting for prey will be successful only when the herbivores come very close (less than 10 to 100 m) to the carnivorous mammals (Elliott et al. 1977). Then, the searching (or predation) efficiency will be correlated with the moving speed of herbivorous mammals, which may be very slow because grass is ample and herbivorous mammals do not need to move fast while foraging unless they have eaten all the grass around them. When the daily movement of herbivores is assumed to be 200 (m/day), and lions will prey successfully with a probability of $P_{hc} = 0.25$ only when herbivorous mammals approach them from the front at a distance of less than 30 m, then the searching volume (area) of lions is $6 \times 10^3$ (m$^3$(m$^2$)/day). Because an individual lion has a volume of 0.2 (m$^3$), S is estimated as S=$3.0 \times 10^4$ (/day). This value may seem rather small, but it can be shown to be reasonable based on the observations that a lion should eat 12 (kg/day) of meat during the wet season (180 days/year), when the population density of large mammals is 100 (individuals/km$^2$) = 0.014 (kg/m$^2$) (as in the Serengeti) (Coe et al. 1976; Packer et al. 2005). The estimation made from these observations, $S = 1.6 \times 10^4$ (/day), is on the same order of magnitude as the above estimation of $S = 3.0 \times 10^4$ (/day). When S=$3.0 \times 10^4$ (/day) is adopted, the herbivore biomass density **h** is calculated from (3)' as **h**=$1.0 \times 10^{-3}$ (kg protein/m$^2$(m$^3$)) = 1.0 (g protein /m$^2$) = 5.0 (g fresh biomass/m$^2$) = 7142 (kg fresh apparent biomass of herbivores including grass inside body /km$^2$) = 35.714 (wildebeest (and/or zebra) individuals/km$^2$) = 527,246 (wildebeests and zebras in Serengeti National Park (14,763 km$^2$)) (Table 3). Then the annual consumption rate of plant leaves is estimated to be 32.6%. Further, the carnivore biomass density **c** is calculated from (4)' as **c**=$5.33 \times 10^{-6}$ (kg protein/m$^2$(m$^3$))=5.33 (mg protein/m$^2$) = 26.6(mg fresh biomass/m$^2$) = 26 (kg fresh biomass/km$^2$)=0.133(lion/km$^2$)= 1,963 (lions in Serengeti) (Table 3).

In regard to the condition of satiation of herbivores, it is apparent that the inequality (11)' is realized because, in the present case, $G_h$ is very small and

$$\frac{\{n_c(1-\alpha_c)+2\alpha_c n_h\}P_{cc}G_h}{2\alpha_c P_{hc}n_h} + \frac{n_c d_c}{\alpha_c n_h} \cong \frac{n_c d_c}{\alpha_c n_h}$$

holds.

Also, it is apparent that

$$e_c > \frac{n_c d_c}{\alpha_c n_h}$$

always holds, because if it does not hold, then carnivores cannot grow.

As a result, (11)' always holds and the satiation of herbivores will not happen.

The estimated **h** fit well with several empirical observations about biomass and/or the population of large herbivores in the Serengeti, which is 8,352-30,481 (kg /km$^2$) for large herbivores and 1,500,000-1,800,000 for wildebeests and zebras in Serengeti National Park (Coe et al. 1976; Packer et al. 2005) (Table 3). Although the biomass data of herbivorous mammals from empirical observations in the Serengeti appears to be approximately 2-3 times higher than the value calculated from the model, this can be partly explained by the fact that large carnivores such as wildebeests move out of the Serengeti in the dry season, while lions stay. Therefore the annual mean population and biomass of herbivores may be roughly half of these observed values, which would be much closer to the values predicted by the present model. Then the annual consumption of fresh leaf mass is calculated as 7142 (mg/m$^2$) × 0.1 × 365 (days) = 260.6 (g/m$^2$ year). Since we assumed 400 (g/m$^2$) leaves exist, the estimated annual consumption rate of leaves by herbivore mammals is 65.2%. This value shows very good agreement with the empirical observation of an average annual consumption rate of 66% (minimum 15% to maximum 94%) in grassland ecosystems in Serengeti National Park, Tanzania and Masai Mara Game Reserve, Kenya (McNaughton 1985) (Table 3). The estimated **c** from the model fits very well with the empirical observations about biomass and/or the populations of lions that make up the largest portion of the carnivore biomass in the Serengeti; The population of lions observed in the Serengeti is 2,500 - 3,000 individuals with a population density of 0.120 - 0.169 (lions/km$^2$) (Wildt et al. 1987; Packer et al. 2005) (Table 3).

*Realization of a "green world" system*

The present model predicts an annual leaf consumption of ca. 2-6% in systems with caterpillars (ectothermal invertebrates) as herbivores, and ca. 65% in systems with large mammals (endothermal) as herbivores and lions as carnivores. The approximately 2-6% of annual leaf consumption predicted in the former systems with caterpillars (ectothermal invertebrates) as herbivores and carnivorous insects or birds as carnivores is low enough for trees to tolerate and compensate for and also poses no problem for the growth of grasses. In this case it is apparent that the "green world" system defined by the HSS hypothesis, which is a world full of palatable plant material remaining uneaten by herbivores, will be realized (Hairston et al. 1960). On the other hand, the annual leaf consumption of ca. 65% for systems with large mammals as herbivores and lions as carnivores is rather high; some trees may tolerate this while many others may not, but most grasses will tolerate it and compensate for any loss of trees if the ecosystem is fairly productive, as in ecosystems with sufficient temperature, precipitation, and nutrients. (Grassland can tolerate much more severe herbivory. Consider a domestic lawn. Although it undergoes more than three mowings a year, which represents a 300% herbivory rate per year or 300% consumption of an LAI of 4 per year, it remains green.) Therefore, while it is not certain that green forest ecosystems will remain, green grassland will at least be realized unless the productivity of the ecosystem is extremely low. The relationship among the intensity of herbivory predicted from the present model, the productivity of the ecosystems, and the ecosystems and realization of lack of realization of "greening" will be discussed in a later session in the context of Oksanen's theories on food web structure and productivity of ecosystems (Oksanen et al. 1981; Oksanen and Oksanen 2000; Oksanen and Olofsson 2009) and Rosenzweig's and Wollkind's theories on food web systems using isocline-based graphical representation (Rosenzweig and MacArthur 1963; Rosenzweig 1971, 1973; Wollkind 1976).

PREDICTIONS DERIVED FROM THE MODEL

The equations (3) (3)' and (4)(4)' predict a series of correlations between various ecological factors and the biomass density of herbivores and carnivores. In addition, the inequalities (10) and (11) predict under which condition the systems will become unstable. Here I present a series of predictions (correlations) that can be derived from the model.

*Predictions in which factor S are involved*

When the searching efficiency of S is high, both the biomass of herbivores **h** and biomass of carnivores **c** will decrease according to equations (3) and (4). The relation is inversely proportional and written as

$$\mathbf{h}, \mathbf{c} \propto {}^{-1}S \quad (12)$$

This means that when carnivores are efficient, both the biomass of herbivores **h** and carnivores **c** (and total animal biomass density) will decrease. On the other hand, when the carnivores are less efficient, animal biomass will increase.

This relation (12) may partly explain why the biomass of herbivorous mammals in the savannah is much larger than the biomass of herbivorous insects in forests (Table 2 and 3). The reason is likely to be the fact that lions are inefficient predators (S=3.0 × $10^4$ (/day)) that cannot run faster than their prey mammals (Elliott et al., 1977) and therefore should depend on a sit-and-wait strategy, while in forest ecosystems, wasps and birds are efficient predators (S=0.75-1.29 ×$10^7$) that can move faster than their prey (e.g., caterpillars whose motion is very slow) and can search large areas by flying.

The relation (12) may partly explain why the belowground ecosystem is rich in animals. Indeed, sometimes the biomass can be as high as 41 (g fresh mass /$m^2$) for earthworms alone (Phillipson et al. 1978; Neirynck et al. 2000), and if other animals and microorganisms in forest soil are included, the biomass would be much higher than that of the aboveground terrestrial ecosystem in a forest scarce in animals with biomasses less than 500 (mg fresh mass /$m^2$) (Schowalter et al. 1981; Meyer et al. 2005). Because the visibility and movement are very much restricted in a belowground system, the searching ability S will be very small in soils. The very small S in the belowground system will make all **h**, **c** and total animal biomass (**h** + **c**) values much larger in belowground ecosystems than in aboveground ecosystems. If S is several orders of magnitude smaller belowground, this will explain and predict **h** and **c** values that are several orders of magnitude higher in belowground than in aboveground ecosystems. Since it is not certain whether the food web structure of belowground ecosystems can be regarded as a 3 trophic-level system (dead leaves and humus as primary materials, microorganisms such as bacteria as primary consumers, and amoeba and earthworms as secondary consumers preying on bacteria), or as a 2 trophic-level system (dead leaves and humus as primary materials, with omnivores such as earthworms prevailing), it is not clear whether the present 3 trophic-level model can be applied to the explanation of the difference between aboveground ecosystems and belowground ecosystems as above, and thus future investigation of this point will be needed. Nevertheless, it is certain from the equations (3) and (4) that the scarcity of animals in the aboveground terrestrial ecosystem can be reasonably explained in part by the large S value compared to $e_h$, which is several orders of magnitude smaller than S.

Since the inequalities (10) and (11) do not contain S, the searching efficiency S does not affect the stability (condition of satiation of carnivores) of the ecosystem.

*Effects of $n_p$ (nutritive value of plants or plant material), $\alpha_h$ (absorption ratio of herbivores feeding on plant) and $e_h$ (speed of plant consumption by herbivores) on **h** (biomass of herbivores), **c** (biomass of carnivores) and the annual consumption rate of plants*

It is clear from equations (3) and (4) that both **h** and **c** increase when $n_p$, $\alpha_h$, and $e_h$ increase. Although both **h** and **c** increase in less than direct proportion to $n_p$, $\alpha_h$, and $e_h$, **c** is closer to direct proportion to $n_p$, $\alpha_h$, and $e_h$ than **h**, whereas **h** becomes close to direct proportion to $n_p$, $\alpha_h$, and $e_h$ when $d_c$ is small. These relationships can explain the following phenomena.

a. Plant-rich ecosystems in the terrestrial world versus animal-rich ecosystems in the aquatic (marine) world

These relationships may explain, at least in part, why terrestrial ecosystems are plant-rich but have small biomasses of herbivores and carnivores, which results in smaller annual consumption rates of plants and realizes a "green world" system, while the aquatic ecosystems are animal-rich with large annual consumption rates of phytoplankton. The great difference between terrestrial and aquatic ecosystems lies in the nutritive value of plants $n_p$; while the protein concentration of terrestrial plants is ca. 2 (% of fresh plant material), that of phytoplankton (diatom, *Thalassiosira pseudonana*) is very high and reaches 25.7 (% of wet plant material) (Harrison et al. 1990), which is as high as that of the flesh of animals. In the present model, this difference in nutritive values (2% vs. 25%) predicts a roughly 6-fold difference in herbivore biomass (when ectotherms are expected for both herbivores and carnivores), with the higher biomass consuming 6 times the volume of 12 time protein-rich plant material. Therefore, a ca. 70-fold increase in the biomass of plant protein per unit volume ecosystem is expected to be consumed in aquatic (marine) ecosystem than in terrestrial ecosystems, and this calculation fit well with the empirical observation that the average percentage of primary production consumed by herbivores, turnover rate (day$^{-1}$), and consumption by herbivores (gCm$^{-2}$day$^{-1}$) are 10-100 times higher in marine ecosystems (phytoplankton) than in forests (leaves of trees), even though the net primary production (gCm$^{-2}$day$^{-1}$) is almost the same between marine and forest ecosystems (Cebrian 1999). Since it is not always the case that the food web structure of a particular aquatic ecosystem can be regarded as a 3 trophic-level system (e.g., phytoplankton, zooplankton, sprat (zooplankter)), and the food-web structure of many aquatic ecosystems are more adequately regarded as 4 trophic-level systems (e.g., phytoplankton, zooplankton, sprat, cod (piscivore)) (Casini et al. 2008), it is not clear whether the present 3 trophic-level mathematical food-web model can be applied to the explanation of the difference between aboveground terrestrial ecosystems and aquatic ecosystems as above. Casini et al. 2008 reported that in the Baltic Sea, the biomass of zooplankton is lower and that of phytoplankton higher when the three trophic-level ecosystem (phytoplankton-zooplankton-sprat system) is realized than when the four trophic-level ecosystem (phytoplankton-zooplankton-sprat-cod system) is realized (Casini et al. 2008). However, even when the three trophic-level system prevails in the Baltic Sea, the biomass of zooplankton (at least 6,000 mg fresh mass/m$^2$ ecosystem of 0-20 m depth range, calculated from Casini et al. 2008) is still much higher than the biomass of herbivores in many forest systems 31.2-360 mg fresh mass/m$^2$ ecosystem, Fig. 2), even though the plant biomass in the three trophic-level system in the Baltic Sea has been estimated to have a chlorophyll concentration of 60 (mg chlorophyll/m$^2$ ecosystem of 0-20 m depth range) (calculated from Casini et al. 2008), which is much lower than that in forest canopies (ca. 400 mg chlorophyll/m$^2$ forest canopy) (Le Marie et al. 2008). This comparison indicated that the aquatic three trophic-level food web systems are much more animal-rich than the terrestrial three trophic-level food web systems, even though plant biomass is more abundant in the terrestrial three trophic-level system, and the present mathematical food web model based on three trophic-level food web systems reasonably explains this difference partly from the fact that $n_p$ is much higher in aquatic systems than in terrestrial systems. Further expansion of the model to four trophic-level food web systems, however, will be needed to discuss the aquatic systems more generally and more in detail. Nevertheless, it is certain from equations (3) and (4) that the scarcity of animals in the aboveground terrestrial ecosystem can be reasonably explained, at least in part, by the approximately ten-fold smaller value of $n_p$ compared to $n_h$ and $n_c$.

b. Relationship between the nutritive values of plants and abundance of herbivores (intensity of herbivory)

It has been repeatedly reported that the biomass of herbivores and intensity of herbivory (plant damage) is high when plants are nutritive with high levels of protein and other nutrients (Siemann 1998; Haddad et al. 2000; Throop and Lerdau 2004). This trend has been taken for granted, but the reason and mechanism for the trend have not been clear. Equation (3) of the present model clearly predicts the observed trend, and thus the present model provides a reasonable explanation for the trend: a greater amount of nutritive plant material supports a greater amount of biomass of both herbivores and carnivores, and *vice versa*. It is also clear form equation (3) that a large value of $\alpha_h$ (large digestibility of plants by herbivores) has a similar effect as a large value of $n_p$ (highly nutritive plants).

c. The ecological mode of function of anti-nutritive defenses such as tannin and digestive inhibitors

Anti-nutritive defenses (low nutritive value, reduction of digestibility and consumption speed by tannin and inhibitors of nutritive enzymes), which mean low $n_p$, $\alpha_h$, and $e_h$, are the most frequently observed defense mechanisms of plants against herbivores (Felton and Gatehouse 1989; Schoonhoven et al. 2005; Zhu-Salzman et al., 2008). Anti-nutritive defenses (low nutritive value, tannin and phenolics, degradation of nutrients, and inhibitors of digestive enzymes) are usually not lethal, although they significantly slow down the growth speed of herbivores (Karban and Baldwin 1997; Konno et al. 1999, 2009). Looking at such mechanisms more closely, however, it is not immediately clear that what we call the "anti-nutritive defenses of plants" will actually benefit the plants, since the herbivores feeding on plants that have anti-nutritive defenses may stay on the plant longer than those feeding on undefended plants, and therefore the herbivores on plants with anti-nutritive defense may ultimately consume a larger amount of plant matter—i.e., an amount sufficient to fulfill their nutritive requirements—than those feeding on undefended plants, and in this case, what we call "anti-nutritive defenses" may result in greater damage to plants and thus cannot properly be considered "defenses" at all. The present model (i.e., equation (3)), however, clearly shows that as a result of the nutritive defenses of plants (i.e., low $n_p$, $e_h$, $\alpha_h$, and $G_h$), **h** will decrease, which will directly result in a lower rate of damage to the plants, and this is the true mechanism by which nutritive defenses function as plant defenses that decrease plant damage. It could be said from equation (3) that being less nutritive (either through a reduction of nutrients or digestibility) is a very effective mode of plant defense via a reduction in herbivore biomass and thereby a reduction in plant damage by herbivores in a natural ecosystem through the function of the food web. From the above, it is reasonable to conclude that the beneficial role of anti-nutritive defenses for plants, such as low plant quality, tannin, phenolics, factors that degrade nutrients, and inhibitors of digestive enzymes, can be understood only at the community level by taking the food web structure into account. The effects of inducible nutritive defenses will be discussed in a later section.

*Implications for plant–herbivore interaction: History, evolution, and coevolution*

In the more than fifty years since the study of coevolution was published (Ehrich and Raven 1964), there have been many experimental studies on the defense mechanisms of plants against herbivores and the adaptive mechanisms of herbivorous insects (Harborne 1993). Although there have been arguments about whether such plant-herbivore interactions really constitute coevolution (i.e., a biological arms race) or just an insect adaptation to plant defense traits, it is clear that for most plant species with so-called defenses, there are specialist herbivores that have developed perfect adaptations, either physiologically or behaviorally, to the strong and often lethal defenses of plants, and that can feed and grow on the plants as if there are no defenses at all (Berenbaum 1978; Dussourd and Denno 1991; Holzinger and Wink 1996; Konno et al 1997, 1999, 2004, 2006, 2009, 2010; Hirayama et al. 2007; Agrawal and Konno 2009; Konno 2011). One may expect from this situation that the specialist herbivores will achieve a breakout or become sufficiently abundant to defoliate all the leaves in a particular ecosystem, resulting in the "browning" of the "green-world" system, now that the specialist herbivores have completely overcome the plant defenses. Our model, however, predicts that such an outbreak will never happen, because even if the specialist herbivores overcome the plant defenses, still the nutritive values of the plants will be much lower than those of the animals. Therefore, even in the case of specialist herbivores that have overcome the plant defenses, their total biomass will be small (a few hundred mg/m$^2$ at the most) and the annual consumption ratio of plants by herbivores will also remain small (a few percent). It is likely that the nutritive value of plants has been low in the past and will be low in the future compared to that of animals over the hundreds of millions of years of plant-insect interaction, and therefore it seems reasonable to expect a consistently low level of herbivory and consistent "green world" system in terrestrial ecosystems from hundreds of millions of years in the past to as many years into the future irrespective of what is called the coevolution of plant defense and insect adaptation, as long as the nutritive value of plants is lower than that of animals and efficient predators exist. In fact, the fossil records of leaf-feeding herbivory (Smith and Nufio 2004; Labandeira and Allen 2007) show that in the Permian forests, the area of leaves damaged by herbivory is 0.25-3.30%, and in Eocene forests, the area of leaves damaged by herbivory is 1.4-2.5%, and neither range is very different from that (0-14%) obtained from modern-day forests (Cebrian 1999), indicating that the level of herbivory and herbivore biomass have been consistently low throughout the history of life on earth. It is likely that this consistency comes from the structure of the food web presented in the present model.

*Effects of species richness on the biomass of animals **h**, **c**, and the intensity of herbivory*

As a result of evolution (speciation and extinction), the species numbers of plants, herbivores and carnivores have varied widely, and even at present there is a great variation of species numbers among ecosystems. It seems, however, that the intensity of herbivory has stayed rather constant among various ecosystems whether or not these ecosystems are tropical, temperate, present, or past (Table 2) (Coley and Barone, 1996; Cebrian 1999; Smith and Nufio 2004; Labandeira and Allen 2007), even though the species number may differ greatly among the systems. Our model predicts that as long as most carnivores are polyphagous (i.e., birds, carnivorous wasps, lizards, lions, etc.), $P_{hc}$ and $P_{cc}$ will remain constantly high and will not be affected by the number of species in the ecosystem. Then, **h**, **c** and the intensity of herbivory (annual consumption rate of plants) will remain constantly low independent of the species number. In short, our model predicts that **h**, **c** and the intensity of herbivory are not largely affected by species number as long as the carnivores are polyphagous and not specialized in particular prey. In contrast, the model predicts that when carnivores are mostly specialists and feed on limited herbivore and carnivore species such as parasitoids, then $P_{hc}$ and $P_{cc}$ (especially $P_{hc}$) will become smaller as the species number (richness) increases, and consequently, **h**, **c**, and the intensity of herbivory will increase as the species number within the ecosystem increases. Further study is needed to assess the effects of generalist and specialist carnivores, but the fact that **h**, **c**, and he intensity of herbivory stay rather constant among ecosystems (Table 2) suggests that generalist carnivores are prevailing and have had a larger influence throughout the terrestrial world and throughout the history of evolution than specialist carnivores.

*Effect of the relative growth rate of herbivores $G_h$ on **h**, **c** and the **h/c** ratio; The food web structure is not always pyramidal with a small carnivore biomass relative to the herbivore biomass if $G_h$ is much larger than $d_c$, such as in a food web structure in which insects are both herbivores and carnivores.*

The equations (3)' and (4)' which describe **h** and **c** using the factor

$G_h =: \frac{\alpha_h e_h n_p - n_h d_h}{n_h}$ (/ day) (relative (inner) growth rate of the herbivore set which feeds freely and continuously on plants under ideal conditions)

$h = \frac{n_c\{(1-\alpha_c)P_{cc}G_h + 2d_c P_{hc}\}}{2\alpha_c P_{hc}^2 S}$ (kg/m³) (3)'

$$c = \frac{n_c G_h}{P_{hc} S} \text{ (kg/m}^3\text{) (4)'}$$

clearly reveal the following.

First, the biomass of carnivores **c** is directly proportional to $G_h$ (the relative inner growth rate of herbivores).

Second, the biomass of herbivore **h** increases when $G_h$ increases. Although **h** increases in less than direct proportion to $G_h$, the relation becomes closer to direct proportion when $d_c$ is small (i.e., when carnivores are ectothermal, such as in the case of insects and reptiles), which means the effects of the nutritive value of plants or nutritive defenses are more prominent when carnivores are ectothermal.

The herbivore/carnivore ratio, h/c, is expressed as follows using (3)' and (4)'.

$$\frac{h}{c} = \frac{(1-\alpha_c) P_{cc}}{2\alpha_c P_{hc}} + \frac{d_c}{\alpha_c G_h} \quad (12)$$

In an ordinary case when $P_{hc} > P_{cc}$ and $\alpha_c$ is nearly 1, the first term of the right side will be much smaller than 1 (e.g., $\frac{(1-\alpha_c) P_{cc}}{2\alpha_c P_{hc}} = 0.025$ when $P_{hc}$=0.25, $P_{cc}$=0.05, $\alpha_c$ = 0.8). The second term of the right side of (12) differs greatly among ecosystems, especially between the ecosystems in which the herbivores are endothermal mammals (such as wildebeests on the African savannah) and those in which the herbivores are ectothermal insects (such as caterpillars in forests). When the herbivores are mammals, the relative inner growth rate of the herbivore set per day $G_h$ (e.g., 0.0002 is assumed in wildebeests; see the lion-wildebeest system above) is much smaller than the relative decrease of biomass from respiration per day $d_c$ (e.g., 0.03 is assumed for lions). Then the second term on the right side becomes very large and as a result **h/c**>>1 is realized (e.g., **h/c**= $5.05 \times 10^2$ (real fresh mass (excluding the gut contents of herbivores) basis or protein basis) is estimated from the model in the lion-wildebeest system while **h/c** = $3.50$-$5.04 \times 10^2$ (real fresh mass basis or protein basis) is empirically observed in Serengeti National Park) (Coe et al. 1976; Wildt et al. 1987; Packer et al. 2005) and shows a very good agreement with the predicted value. This result fits the commonsense understanding that carnivores are much scarcer than herbivores in the food web and that the food web structure can be illustrated as a pyramid with the carnivores at the top. The model predicts, however, that the situation changes when ectothermal animals such as arthropods are herbivores. When caterpillars are herbivores, the inner growth rate of the herbivore set per day $G_h$ (e.g., 0.296 is assumed for caterpillars eating leaves) is often greater than the $d_c$ of insect carnivores (e.g., 0.006 is assumed for insects) or birds (e.g., 0.03 is assumed). Then the second term of the right hand is small and **h**/**c**≤1 is realized; e.g., the present model estimates **h**/**c** values of 0.051 and 0.153 (real fresh mass or protein basis) in the insect carnivore–insect herbivore system and bird-caterpillar system, respectively, while **h**/**c** = 0.23-0.6 in chewing herbivorous arthropod/arthropod carnivores in the forests of North Carolina (calculated in real fresh mass basis or protein basis where the gut contents of herbivores are excluded from the estimation) (Schowalter et al. 1981). In the food web system where the herbivores are insects and $G_h$ is larger than $d_c$, the present model predicts that carnivores can be even more abundant than herbivores, and the prediction fits well with empirical observations. In this case, the common sense understanding in terms of the food web structure that carnivores are less abundant than herbivores is not always true, and it is not appropriate to illustrate the food web as a triangle or pyramid with the carnivores at the top (i.e. it is more appropriate to illustrate such a food web as an inverted pyramid in this case). It is likely that the high **h**/**c** ratio in the wildebeest-lion system in the African Savannah is caused by a low $G_h$ of the mammalian herbivore community and subsequent high $d_c/G_h$ ratio. Similarly, the low **h**/**c** ratio in the caterpillar – insect carnivore (or bird) system in forests is caused by the high $G_h$ of caterpillars, and subsequent low $d_c/G_h$ ratio.

*Conditions that favor outbreak of herbivores: Low $P_{hc}$, high $P_{cc}$, (Low $P_{hc}/P_{cc}$ ratio), high $n_p$, high $d_c$, high $G_h$.*

Whether or not a stable equilibrium is realized or an outbreak of herbivores through satiation occurs can be judged by inequalities (10) and (11). Under ordinary conditions in the forest ecosystem, such as those set above for the wasp (bug)-caterpillar system or bird-caterpillar system, the right side of inequalities (10) and (11) is 0.07356 and 0.1036, and because these values seems to be smaller that $e_c$, which is expected to be 0.15 – 1.0, the system is stable. Under these conditions, $P_{cc}$=0.05, $P_{hc}$=0.25 and $P_{cc}/P_{hc}$=0.2. The situation changes, however, when $P_{hc}$ is smaller and especially when $P_{hc}$ is even smaller than $P_{cc}$, and when the $P_{cc}/P_{hc}$ ratio is high. For example, when $P_{cc}$=0.05, $P_{hc}$=0.025 $P_{cc}/P_{hc}$=2 then the right side of the inequality is 0.669 and 0.699, for the wasp (bug)-caterpillar system and the bird-caterpillar system, respectively, and these values are almost the upper limit for $e_c$. Therefore, if $P_{hc}$ is much smaller than 0.025, carnivores can no longer control the population (biomass) of herbivores and outbreak is likely to occur. Even if the $e_c$ is larger than these values and the outbreak does not occur, very high **h** is expected from equation (3) and the annual consumption rates of leaves are expected to be 6.218%, 21.98%, and 14.81% for the wasp-caterpillar system, the carnivorous bug-herbivorous beetle system, and the bird-caterpillar system, respectively. Such extreme conditions with low $P_{hc}$ typically occur when the caterpillars are toxic or hazardous and avoided by most predators, such as in the case of the Danainae and Troidini butterfly larvae, which sequester toxic compounds from their host plants, Zygaenidae larvae and Chrysomelidae beetle larvae and adults, which produce toxins or defense chemicals by themselves, or the hairy caterpillars of Lymantriidae, Lasiocampidae, and Limacodidae species, which have allergenic hairs (Harborne 1993; Nishida 2002). These insects very often achieve a breakout and defoliate all the leaves on plants. The $P_{hc}$ will also be small if the herbivore insects can move fast and evade their predators. Locusts and many other orthopteran insects, planthoppers (Hemiptera), and fleebeetles that can jump also tend to exhibit breakouts, and this can be explained by their low $P_{hc}$, which is due to the evasions made possible by their jumping ability. Apart from low $P_{hc}$, conditions such as high $P_{cc}$, high $n_p$, high $e_h$, high $G_h$, and high $d_c$ will contribute to the likelihood of an outbreak of herbivores, which means that the likelihood an herbivore outbreak increases when the nutritive value of plant $n_p$ is high, the relative inner growth rate of herbivores is high, the carnivores in endothermal ($d_c$ is high), and the possibility of intraguild predation $P_{cc}$ is high. Even if the possibility of intraguild predation is not high enough to cause the outbreak of herbivores predicted by inequality (10), equation (3) predicts that the biomass of herbivores and intensity of herbivory increase as the possibility of intraguild predation becomes larger, which agrees with the empirical observations on the effect of intraguild predation (Rosenheim et al. 1993), but our model predicts the trend even without assuming any structure inside the carnivore set (e.g., top predators and low level predators).

*Global equilibrium and local variation in conditions; the reason why inducible defenses are effective and the reason why young leaves in the tropical forest are severely damaged.*

Thus far in the discussion, the present model has assumed a global uniformity of conditions ($n_p$, $G_h$, etc.) in a large area, and has successfully calculated the global equilibrium in biomass of herbivores **h**, and equilibrium in biomass of carnivores **c**. The herbivores stay at equilibrium because the relative inner growth rate of herbivores $G_h = \frac{\alpha_h e_h n_p - n_h d_h}{n_h}$ balances the relative predation rate of herbivores $\frac{P_{hc}Sc}{n_c}$. When global equilibrium $c = \frac{n_c(\alpha_h e_h n_p - n_h d_h)}{P_{hc} n_h S}$ is realized, then the relative predation rate of herbivores is $\frac{\alpha_h e_h n_p - n_h d_h}{n_h} = G_h$ and balances the relative inner growth rate of herbivores, and the population stays at equilibrium. Global uniformity in the ecosystem, however, is not always realized and there are several types of heterogeneity. Next, I will consider what might be expected from the present model in the case that homogeneity in an environment is not expected.

First, I will discuss the ecological consequence of inducible plant defenses against herbivorous insects, such as are very widely found in many plant species (Karban and Baldwin, 1997). Many defensive substances or traits are induced when plants are damaged by herbivores, which decreases the performance of herbivores feeding on them (Karban and Baldwin, 1997). What is the effect of the local defense and/or inducible defense (inducible defenses are mostly local) of plants on the biomass density of herbivores? Here, by "local" I mean that the area of plant defense is more local (sporadic) than the searching range of an individual carnivore, that a carnivore cannot perceive the existence of the local defense, and that the population density of carnivores remains unchanged at the global equilibrium $c$. When the relative inner growth rate of herbivores on locally defense-induced plants becomes $\gamma G_h$ ($0 \leq \gamma < 1$) because of a plant defense or a decrease in nutritive value, and when the relative predation rate of herbivores by predation remains at the global rate $\frac{\alpha_h e_h n_p - n_h d_h}{n_h} = G_h$, then the local population density of herbivores at the locally defended or defense-induced plant decreases at the rate of $(1-\gamma)G_h$ (/ day) or $\frac{(1-\gamma)(\alpha_h e_h n_p - n_h d_h)}{n_h}$ (/day). For example, when global conditions attain (e.g., on undefended plants) such that caterpillars can realize a relative inner growth rate of 0.30 (the biomass doubles once in three days), and when on locally defended and/or defense-induced plants caterpillars can realize a relative inner growth rate that is half (0.15 when $\gamma = 0.5$) that of undefended plant, then, the relative decrease rate of the biomass of herbivores will be 0.30-0.15=0.15 (/day), which means that the biomass density of herbivores initially hatched from eggs at day 0 will be reduced by half by day 6. If caterpillars remain on the plants feeding for 24 days after hatching and before pupation in the case of undefended plants (or for 48 days on locally defended plants), then the expected damage inflicted on the locally defended/defense-induced plants (calculated as the leaf area based on integral calculus) will be approximately 2.8 times smaller than that of the undefended plant in the surrounding ecosystem even when same amount (number) of eggs are laid on both types of plants, as determined by calculating the integration of the exponential function, and most of the damage will be done by young caterpillars. Similarly, if $\gamma = 0.1$ locally, then the damaged area will be 7.6 times smaller, locally. On the other hand, when the $G_h$ globally decreases in a large area in an ecosystem at $\gamma = 0.5$ and 0.1 under the conditions set in the caterpillar–insect herbivore system (see above), equations (3) and (3)' suggest that the amount (area) of damage would be 1.3 times and 1.9 times smaller, respectively, and in the caterpillar-bird system, the amount (area) of damage will be suggested to decrease by only 1.1 times and 1.2 times. These results indicate that if the plant is sporadically low in nutritive value or is sporadically defended compared to the surrounding (undefended) plants in the ecosystem, the plant will remain almost entirely undamaged with a very low local biomass of herbivores. These results further indicate that inducible defenses are effective not only because of the defense activities *per se*, but also or even more because they are local.

Under the opposite conditions, in which the plant has a locally high nutritive value that supports the relative inner growth rate of $\gamma G_h$ ($1<\gamma$), the plant will be severely damaged.

For example, when the relative inner growth rate of caterpillars on ordinary leaves is 0.3 (doubling in 3 days, and with the caterpillars staying on the leaves for 24 days after hatching) and

γ=1.6, and the relative inner growth rate on a plant with locally high nutritive value is 0.5 (doubling in 2 days, and with the caterpillars staying on the leaves for 14 days) (as new leaves coming out from the leaf bud sporadically or a highly nutritive cultivar of plant or a highly fertilized plant individual planted solely in field among many low-nutritive plant individuals), the increase rate of the herbivore biomass is 0.2 (/day), and integral calculus reveals that a 2.9-fold higher amount (area) of lead damage is expected for the sporadically high nutritive plants than in ordinary leaves in the surrounding environment, even when the same amount (number) of eggs are laid on both types of plants. If the herbivores preferably lay eggs on young leaves, the damage will be much greater unless the carnivores perceive the high density of herbivores. In contrast, when the nutritive value of the leaves increases globally at a rate of γ = 1.6 in the caterpillar–insect herbivore system and caterpillar–bird system, only 1.3-fold and 1.1-fold increases in the damage area are expected, respectively. This may explain why young leaves that sporadically come out will suffer high rates of damage (the area of leaves consumed as young leaves is 7.6% of the total leaf area, and the daily consumption rate of young leaves is 0.71%/day) compared to mature leaves in tropical forests (the area of leaves consumed as mature leaves is 2.4% of the total leaf area, and the daily consumption rate of young leaves is 0.03%/day), and compared to young leaves that come out globally (throughout the forest) in the temperate deciduous forests (the area of leaves consumed as young leaves is 1.9% of the total leaf area while those consumed as mature leaves is 5.2%) (Coley and Barone, 1996).

Similarly, it is also expected that the intensity of damage will differ depending on the conditions of the surrounding global environment (i.e., the nutritive value, defense activity and corresponding $G_h$ of the surrounding forests). The same individual plant with the same nutritive value and $G_h$ for herbivores will suffer greater leaf damage when placed among plants with a low nutritive value, than when placed among plants with a high nutritive value.

From the above discussion, it is now reasonable to consider equation (4) (and (4)') in the present model as predictors of the global field (or global rate) of predation pressure in a particular ecosystem.

*Connection of the present parameterized mathematical food web model with the previous food web models in terms of graphical representation based on isocline, stability, and ecosystem productivity*

Although the preceding influential mathematical food web theories and models, many of which adopted graphical representation based on isoclines, have clarified the qualitative traits of the food web structure and the conditions that lead to stable equilibrium of the food web, and have predicted the existence of limit conditions at which the food web structure shifts from one status to another as the ecological factors change (Rosenzweig and MacArthur 1963; Rosenzweig 1971; Rosenzweig 1973; Wolkind 1976; Oksanen et al. 1981; Oksanen and Oksanen 2000; Oksanen and Olofsson 2009), these models did not predict the limit conditions with practical physiological units, and therefore it has been impossible to predict the current status of our ecosystem. For example, Oksanen and his colleagues predicted and demonstrated that as the productivity of the ecosystem (or plant growth speed) decreases, the food web structure shifts from a three trophic-level structure with carnivores, herbivores, and plants that realizes a "green world" system to a two trophic-level structure with carnivores and herbivores or to a one trophic-level system consisting of plants alone, both of which are "less green," and similarly forest ecosystems tend to shift to grass land systems as the productivity decreases (Oksanen et al. 1981; Oksanen and Oksanen 2000; Oksanen and Olofsson 2009), but it has been difficult to predict logically when (at what particular level of productivity) these shifts take place, or the current status of our ecosystem. Similarly, the preceding models predicted that while the handling time of carnivores destabilizes food-web systems, intraguild predation of carnivores stabilizes them (Rosenzweig and MacArthur 1963; Willkind 1976; Oksanen and Oksanen 2000), but it has been hard to predict from these models which force, destabilizing or stabilizing, is significant in a particular ecosystem or under particular conditions. The present parameterized food web model made it possible to predict these points.

In order to understand what the present parameterized model suggests, I drew isocline graphs for the forest ecosystem and savannah ecosystems (Fig. 2). The isocline graph in Fig. 2A represents the model of the forest ecosystem with herbivorous beetles as herbivores and carnivorous bug as carnivores. Carnivore biomass increases in the area right of the carnivore isocline (red line) and decreases in the area left of the carnivore isocline, while herbivore biomass increases in the area below the herbivore isocline (green and brown lines) and decreases in the area above the same lines. In regard to the carnivore isocline line (red line), it stands straight (perpendicularly) up from the horizontal axis at first, but gradually inclines to the right as it rises. This inclination comes from the intraguild predation of carnivores. In regard to the herbivore isocline (green and brown lines), the line stays constant (flat) when the herbivore biomass is low (but not very low) and carnivores are not satiated by herbivores. In this phase, there is no effect coming from the limitation of digestive speed (capacity) or handling time of carnivores. When the herbivore biomass surpasses a particular point, carnivores start to be satiated by herbivores, and the line ascends as it goes right (and the relation is proportional). While the biomass of herbivores and intensity of herbivory are low, both trees and grass can tolerate and compensate for the herbivory and plant biomass, allowing them to remain stable over time without any rapid drops. As the biomass of herbivores and intensity of herbivory increase and surpass a particular point, the trees first become unable to tolerate herbivory (see brown line) and then the tree biomass drops rapidly but the grass biomass stays constant (see green line). At this point, a shift from forest to grassland is expected, as Oksanen has predicted and observed (Oksanen et al. 1981; Oksanen and Oksanen 2000; Oksanen and Olofsson 2009). But as the herbivore biomass further increases, grass, too, becomes unable to tolerate the herbivory, and the biomass of grass decreases rapidly. In this case, the stable (convergent) "green world" system (or green grassland) composed of three trophic levels will collapse and will shift to a less green ecosystem composed of only two trophic levels (herbivores and plants) or a single trophic level (plants only), as predicted and observed by Oksanen (Oksanen et al. 1981; Oksanen and Oksanen 2000; Oksanen and Olofsson 2009). Considering these facts together, there are four important points in the herbivore isocline, the point of equilibrium, the point where carnivores begin to be satiated by herbivores, the point where trees can no longer tolerate herbivory, and the point where grass can no longer tolerate herbivory. Since the previous non-parameterized models could not predict the order of these four points in a particular ecosystem in a logical manner, it has been impossible to predict the stability of the ecosystem; on the other hand, in the case of a stable ecosystem, it has been impossible to predict which type of ecosystem will be realized, a "green forest" or "green grassland" system, both of which consist of three trophic levels, or a less green ecosystem composed only of one or two trophic levels. The present parameterized mathematical food-web model succeeded in predicting the order (Fig. 2). Equation (3) in our model predicted that under an equilibrium condition in the forest ecosystem with carnivorous bugs as carnivores and herbivorous beetles as herbivores, the intensity of herbivory $I_h$ is 4 (% to LAI of 4/year) or 16 (g wet mass leaves/m$^2$ ecosystem year) (Table 2, Fig. 2A). Since we have already clarified that the right side of inequality (11) is 0.07365, and is probably ten times smaller than the $e_c$ in the left side, we can state that carnivores under the equilibrium condition account for only 1/10 of the amount of prey that carnivores can ideally consume and digest. Therefore the point of $I_h$ at which carnivores will be satiated by herbivores is much (ten times or more) higher than the point of $I_h$ at the equilibrium, and the value is likely to be 40 (% to LAI of 4) or more. It is reasonable to assume that in a productive ecosystem with warm temperature, ample rainfall, and fertile soil, the trees will be able to tolerate and compensate for an $I_h$ of up to 50 (% to LAI of 4/year) or 200 (g wet mass leaves/m$^2$ ecosystem year), and grass will be able to tolerate an $I_h$ of up to 300 (% to LAI of 4/year) or 1,200 (g wet mass leaves/m$^2$ ecosystem year). Therefore, $I_h$ under the equilibrium condition is smaller than $I_h$ at the point of carnivore satiation and the equilibrium point is in the phase where the isocline of herbivores is constant and flat, which means that the destabilization arising from the handling effect or the satiation of carnivores does not exist at the equilibrium point and only the stabilizing effect from intraguild predation of carnivores exists there. This indicates that the equilibrium point is stable and convergent. Some previous studies have assumed that density dependence caused by a limitation of resources and competition for resources in herbivores is an important stabilizing force for food-web systems (Rosenzweig and MacArthur 1963; Rozenzweig 1973). However, the stabilizing effect derived from resource competition among herbivores does not seem to be significant, because the present model predicted a very small herbivory rate (4% annual consumption) in the forest food-web system and it is not likely that herbivores are competing for limited food resources, and also because the stability of the food-web system around equilibrium is reasonably explained by other factors (i.e., the stabilizing effect from intraguild predation of carnivores). In addition to the stability of the equilibrium, $I_h$ under the equilibrium condition is much smaller than those at the limit points that tree and grass can tolerate. In conclusion, both the stable green forest and stable green grassland ecosystems, which are both "green-world" systems consisting of three trophic levels, can exist, but because trees are more fierce competitors than grass in terms of capturing light, a stable green forest will be realized in this case. As long as the tree productivity is high enough to tolerate and compensate for the herbivory intensity $I_h$ of 4 (% to LAI of 4/year) or 16 (g wet mass leaves/m$^2$ ecosystem), a stable green forest can persist. In fact, in most places on earth where trees can survive and grow, trees are likely to tolerate an annual herbivory of 4%, and forests will likely to prevail in an ecosystem with very low productivity if herbivores are mostly invertebrates, and herbivorous mammals such as deer are absent. In many areas of the world, including mountainous regions of South East Asia and East Asia, evergreen shrub forests consisting of *Ericaceae* plants, including dwarf *Rhododendron* and *Vaccinium*, dwarf pines, etc., which seem unlikely to tolerate severe herbivory, persist nearly up to the snow line where large herbivorous mammals are absent, but herbivorous insects are present.

The isocline graph in Fig. 2B provides information on the savannah ecosystem with wildebeests (or zebras) as herbivores and lions as carnivores. In this case, the present parameterized model predicted higher $I_h$ values of 65 (% to LAI of 4/year) or 260 (g wet mass leaves/m$^2$ ecosystem) at equilibrium in the savannah compared to the $I_h$ values at equilibrium in the forest ecosystem with insect herbivores and carnivores: 4 (% to LAI of 4/year) or 16 (g wet mass leaves/m$^2$ ecosystem). Even in this case, however, inequality (11) showed that the carnivores were not satiated by herbivores (as shown previously in this paper), and the $I_h$ values of 65(% to LAI of 4/year) or 260 (g wet mass leaves/m$^2$ ecosystem) at equilibrium were still lower than those estimated at the point on the herbivore isocline where carnivores start to be satiated by herbivores: $I_h$ = 260(% to LAI of 4/year) or 1,040 (g wet mass leaves/m$^2$ ecosystem). This indicated that the equilibrium point is in the phase in which the isocline of herbivores is constant and flat, and that the equilibrium is stable and convergent with a stabilizing effect on the intraguild predation of carnivores but without the destabilizing effect caused by the satiation of carnivores by herbivores (or handling time). Although the $I_h$ at the equilibrium, 65 (% to LAI of 4/year) or 260 (g wet mass leaves/m$^2$ ecosystem year), was considerably lower than the $I_h$ at the limit condition that grass can tolerate, 300 (% to LAI of 4/year) or 1,200 (g wet mass leaves/m$^2$ ecosystem year), it was slightly higher than the $I_h$ at the limit condition that trees can tolerate, 50 (% to LAI of 4/year) or 200 (g wet mass leaves/m$^2$ ecosystem year). This comparison means that while the stable green forest is vulnerable or may not exist, it is very likely that the stable green grassland with three trophic levels will persist. In the food-web system with lions as carnivores and wildebeests as herbivores, stable green grassland with three trophic levels will be maintained if the productivity of the ecosystem is high enough that grass can tolerate an herbivory intensity of 65(% to LAI of 4/year) or 260 (g wet mass leaves/m$^2$ ecosystem year), which is likely to be fulfilled. Meanwhile, for a forest to exist under the lion-wildebeest system, trees should also be able to tolerate herbivory intensity of 65(% to LAI of 4/year) or 260 (g wet mass leaves/m$^2$ ecosystem year), which is an unlikely condition except in the case of very productive ecosystems. If the productivity of the ecosystem (= growth speed of plant) is low and even grass cannot tolerate an herbivory intensity of 65(% to LAI of 4/year) or 260 (g wet mass leaves/m$^2$ ecosystem year) in the lion-wildebeest system, even stable green grassland with three trophic levels will not be able to survive, and the ecosystem will shift into a far less green ecosystem in which only a small part of the land is covered with vegetation, but since the present model is based on a three trophic-level food web, the model cannot accurately predict what will really happen in such low-productivity ecosystems, which seems to be the limitation of the present model. The relatively large values in limit intensity of herbivory come from the relatively large herbivore biomass in the lion-wildebeest system, which are likely to have resulted from the slower running speed of lions compared to that of wildebeests and consequent low S (searching efficiency) of lions. In northern ecosystems, the wolf-deer system is a common carnivorous mammal-herbivorous mammal system, instead of the lion-wildebeest system seen in the African savannah. Unlike lions, wolves are marathon runner-like chasers, and after a long chase, they catch up with deer with high probability. This means that the S of wolves in a wolf-deer system is larger than the S of lions in a lion-wildebeest system. As a result, the biomass of herbivores and limit intensity of herbivory that trees and grass should tolerate to maintain a stable green forest or grassland, respectively, is expected to be lower in the wolf-deer ecosystem in northern areas than in the lion-wildebeest ecosystem on the African savannah. This difference may explain why wolves can keep the biomass of herbivores fairly low, and why a stable green forest is maintained even in a northern ecosystem where the productivity of the ecosystem and the growth speed of plants are very low (lower than on the savannah) if wolves are present as herbivores (Ripple et al. 2010). Although the limit intensity of herbivory that trees and grass need to maintain stable green forest and grassland, respectively, which is the limit productivity of trees and grass needed to maintain stable green forest and grass land, respectively, all of which can be predicted from our present model, is low in the wolf-deer system, the shift from a stable green forest to a stable grassland ecosystem, and finally to an arid land ecosystem with scarce plant cover consisting of two or fewer trophic levels without carnivores will take place as it goes northward and the productivity of plants decreases, but the details of such a low-productivity ecosystem with less than two trophic levels cannot be explained using the present model based on a three trophic-level food-web system. Although most real terrestrial ecosystems include both insect carnivore–insect herbivore food-web systems (modules) and mammal carnivore–mammal herbivore systems (modules) at the same time, because of the difference in body size in animals between both modules, these modules are less closely connected to each other and somewhat independent (Sinclair et al. 2010), and therefore it is reasonable to consider each module separately as above, and addition of herbivory intensities from these modules can be used to predict which landscape, a green forest, green grassland, or arid land landscape, will appear.

In conclusion, the present parameterized food-web model brought in parameterized realities to food-web studies, which is indispensable to predict which type of ecosystem or landscape will appear.

THE INTUITIVE UNDERSTANDING OF THE MODEL AND WHY THE TERRESTRIAL WORLD IS GREEN

The intuitive understanding of the present mathematical model, which predicts that a "green world" system will necessarily appear in a terrestrial ecosystem, can be obtained by admitting the principle that it is more efficient to search a large area for a rare but highly nutritive food (e.g., animal) rather than to stay and feed on very abundant but low quality food (e.g., plant) existing in the area if the density of the highly nutritive food exceeds the threshold (equilibrium), which is very low, and in this case, the equilibrium condition could be reached only when the feeding efficiency and relative growth rate of carnivores become almost as low as that of herbivores, which condition could be realized only when the total amount of prey (herbivores + carnivores) is low even in the vast area that carnivores can search (S is very large as shown above), and only when carnivores can eat only after a long interval, and this means that under the equilibrium condition, the biomass concentration of herbivores should be very low, which guarantees a "green world" system. The above discussion clearly shows that the satiation of carnivores by herbivores is not likely to happen in equilibrium (because if carnivores are satiated, they can grow much faster than herbivores, because the nutritive value of herbivores is much higher than that of plants, and in this case equilibrium is never obtained). Since the destabilizing effect caused by the satiation of carnivores (or handling time or digestive ability of carnivores) is absent and the stabilizing effect caused by intraguild predation of carnivores exists alone, the equilibrium is stable and convergent. For these reasons, a stable "green world" system emerges in the terrestrial ecosystem.

CONCLUSIONS

1. A novel mathematical model is proposed for a three trophic level food-web system consisting of carnivores, herbivores, and plants that can predict absolute biomass concentrations of herbivores **h** and carnivores **c** with practical physical units as follows.

$$\mathbf{h} = \frac{n_c\{(1-\alpha_c)P_{cc}(\alpha_h e_h n_p - n_h d_h) + 2n_h d_c P_{hc}\}}{2\alpha_c P_{hc}^2 n_h S}$$ (kg protein or stoichiometrically-limiting nutrients/ m$^3$)

$$\mathbf{c} = \frac{n_c(\alpha_h e_h n_p - n_h d_h)}{P_{hc} n_h S}$$ (kg protein or stoichiometrically-limiting nutrients/ m$^3$)

2. The model predicts a stable convergent equilibrium under ordinary ecological conditions in terrestrial ecosystems. A condition that is required for the convergent and stable equilibrium of biomass of herbivores **h** and carnivores **c** in the present model is the existence of intraguild predation of carnivores, which is realized in most ordinary ecosystems. The present model indicates that intraguild predation of carnivores stabilize the food-web system, as is also predicted by previous models. Previous models predicted that handling time destabilizes the food-web system, but our model predicts that at equilibrium under ordinary ecological conditions, the effects of handling time (mostly time needed for digestion, absorption and assimilation) is minimal, because carnivores are not satiated by herbivores and the herbivore isocline line is constant (flat) at the equilibrium. Taken together, this shows that, since only a stabilizing effect from intraguild predation of carnivores exists at the equilibrium and no or only a minimal destabilizing effect from handling time exists, the equilibrium is convergent.

3. The predicted biomass concentrations of herbivores **h** and carnivores **c** showed very good agreement with those obtained from empirical observation in a forest ecosystem where insects are the major herbivores and in savannah ecosystems where mammals are major herbivores; the predicted values and empirical observations were on the same order of magnitude (e.g., the predicted and observed values of **h** and **c** in the forest ecosystem with insects as major herbivores were within the order of magnitude of ca. 100 mg/m$^2$ fresh biomass excluding the gut contents of herbivorous insects, whereas in the savannah ecosystem, where large mammals are both major herbivores and carnivores, the predicted and observed **h** is (ca. 7,000-30,000 mg/m$^2$ fresh biomass) and is much larger than the predicted and observed **c** (24-34 mg/m$^2$ fresh biomass).

4. The model predicted a low annual consumption rate of leaves (ca. 1-6%) in a terrestrial forest ecosystem with insects as herbivores and insects or birds as carnivores, and this low consumption rate guarantees the existence of a "green world" system. The model predicted a higher annual consumption rate of leaves (ca. 65%) on the savannah with lions as carnivores and wildebeests (or zebras) as herbivores, and this consumption rate is probably too high for trees to tolerate and for forest to exist, but still low enough for grass to tolerate and for green grassland to exist.

5. The model predicts a number of relationships, such as: biomass of both carnivore **c** and herbivore **h** is small in ecosystems where carnivores are efficient and can search large areas or where the nutritive value of plants is low and *vice versa* (the herbivore biomass/carnivore biomass) ratio **h/c** is high in ecosystems where the (decreasing speed of carnivore from respiration)/(relative growth rate of herbivore) ratio $d_c/G_h$ is high, as in the case of a savannah where both carnivores and herbivores are mammals, and *vice* *versa* as in forest ecosystems where insects are herbivores; outbreak of herbivores more likely when carnivore prefer to prey on carnivore rather than on herbivores.

6. The model predicts that in an ecosystem where carnivores are polyphagous (generalist), the biomass of herbivores **h** and carnivores **c** and intensity of herbivory will remain low and will not be affected by the species richness of the ecosystem. The model predicts that in an ecosystem where carnivores are specialists and feed on a very limited number of species, the biomass of herbivores **h** and carnivores **c** and intensity of herbivory are high and increase as the species richness of the ecosystem increases. The consistently low biomass of herbivores **h** and carnivores **c**, and the intensity of herbivory through much of the ecosystems suggests that the former condition (generalist carnivores) prevails in most terrestrial ecosystems.

7. The model explains why the aboveground terrestrial ecosystems is plant-rich; the major reasons are that (a) plant is far less nutritive than that animal ($n_p \ll n_h, n_c$), and that (b) the volume (ratio) that carnivore can search per day is much larger than the volume (ratio) that herbivore can eat per day ($S \gg e_h$). In contrast, the belowground ecosystem does not fulfill condition (a) and the aquatic ecosystem does not fulfill condition (b), and these points may potentially and partly contribute to the richness of animal biomass in these ecosystems, especially if these ecosystems are composed of three trophic levels.

8. The model is applicable to investigating the mechanism of nutritive defense by tannins and inhibitors, and the mode of actions of inducible defenses.

9. Although the model treats the ecosystems in a very simplified manner (e.g., no internal structure is assumed within carnivores), it can still make quite reasonable predictions as to the absolute biomass of herbivores and carnivores. It appears that in future, however, more highly developed versions of the present model, including versions with more complicated food-web structures, will be required. Nevertheless, one of the most important achievements of the present study is that it showed how to predict the absolute biomass of trophic levels with practical physical units, the parameterized realities which are essential for understanding and predicting the status of ecosystems, for applying ecological theories to practical problems such as pest management, for the conservation of rare carnivore species, and for quantitatively evaluating the ecological consequences of environment modification by human activities.

ACKNOWLEDGMENTS

I thank Professor Fumiyasu Komaki, Department of Mathematical Informatics, University of Tokyo, for checking the relevance and accuracy of the mathematical model and for invaluable comments that improved the manuscript. I also thank Professor Lauri Oksanen, University of Turku, for reading the manuscript and providing innumerable invaluable comments that helped deepen the discussion. This research was supported by a Research Grant Project from the Ministry of Agriculture, Forestry and Fisheries of Japan (MAFF).

Table 1. Definitions of the variables included in the model

| Variable | Definition | Explanation | Physical Unit |
|---|---|---|---|
| $t$ | Time | | Day |
| $h(t)$ | Total biomass density of herbivore (a set of herbivore) | The biomass density is expressed as kg protein (or kg stoichiometrically limiting nutrients) that exists in unit volume ecosystem (m$^3$) | kg/m$^3$ |
| $c(t)$ | Total biomass density of carnivores (a set of carnivore) | Same as above | kg/m$^3$ |
| **h** | Stable convergent equilibrium of total biomass density of (a set of) herbivore | Same as above | kg/m$^3$ |
| **c** | Stable convergent equilibrium of total biomass density of (a set of) carnivore | Same as above | kg/m$^3$ |
| $V_h(t)$ | Volume ratio of herbivore | Ratio between total volume (m$^3$) of herbivore to the volume (m$^3$) of the particular ecosystem that the herbivore are existing | None (ratio) |
| $V_c(t)$ | Volume ratio of carnivore | Ratio between total volume (m$^3$) of carnivore to the volume (m$^3$) of the particular ecosystem that the carnivore are existing | None (ratio) |
| $F_{out\ p}(t)$ | Outflow of biomass from plant | Biomass that goes out of plant set per unit volume ecosystem per day as a consequence of herbivory | kg/m$^3$day |
| $F_{in\ h}(t)$ | Inflow of biomass into herbivore | Biomass that goes into herbivore set as a consequence of herbivory | kg/m$^3$day |
| $F_{out\ h}(t)$ | Outflow of biomass from herbivore | Biomass that goes out of herbivore set as a consequence of carnivory | kg/m$^3$day |
| $F_{in\ c}(t)$ | Inflow of biomass into carnivore preying on herbivore | Biomass that goes into carnivore set as a consequence of carnivory on herbivore | kg/m$^3$day |
| $F_{out\ cc}(t)$ | Outflow of biomass from carnivore through intraguild predation of carnivores | Biomass that goes out of carnivore set as a consequence of intraguild predation of carnivores | kg/m$^3$day |
| $F_{in\ cc}(t)$ | Inflow of biomass into carnivore through intraguild predation of carnivores | Biomass that goes into carnivore set as a consequence of intraguild predation of carnivores | kg/m$^3$day |
| $F_{out\ c}(t)$ | Net outflow of biomass from carnivore through intraguild predation of carnivores | Net biomass loss from carnivore set as a consequence of intraguild predation of carnivores | kg/m$^3$day |
| $D_h(t)$ | Loss of herbivore biomass in form of respiration, metabolism, and excretion | Biomass (as protein) lost from herbivore set as a result of respiration, metabolism, and excretion per unit volume of ecosystem (m$^3$) per day | kg/m$^3$day |
| $D_c(t)$ | Loss of carnivore biomass in form of respiration, metabolism, and excretion | Biomass (as protein) lost from carnivore set as a result of respiration, metabolism, and excretion per unit volume of ecosystem (m$^3$) per day | kg/m$^3$day |
| $n_p$ | Nutritive value of plant | Mass (kg) of protein (or stoichiometrically limiting nutrients) contained in unit volume (m$^3$) fresh biomaterial | kg/m$^3$ |
| $n_h$ | Nutritive value of herbivore | Same as above | kg/m$^3$ |
| $n_c$ | Nutritive value of carnivore | Same as above | kg/m$^3$ |
| $e_h$ | Feeding efficiency (maximum eating speed) of herbivore | Ratio between volume of plant material that herbivore can eat per day (m$^3$/day) in the ideal condition and volume of herbivore (m$^3$) | /day |
| $e_c$ | Feeding efficiency (maximum eating speed) of carnivore | Ratio between volume of animal material that carnivore can eat per day (m$^3$/day) in the ideal condition and volume of carnivore (m$^3$) | /day |
| $\alpha_h$ | Absorption efficiency of herbivore feeding on plant | Ratio between plant biomass (as protein) absorbed by and incorporated into the body of herbivore and plant biomass (as protein) eaten by herbivore | None (ratio) |
| $\alpha_c$ | Absorption efficiency of carnivore feeding on animal | Ratio between animal biomass (as protein) absorbed by and incorporated into the body of carnivore and animal biomass (as protein) eaten by herbivore | None (ratio) |
| $S$ | Searching efficiency of carnivore | Ratio between volume that carnivores search and volume of carnivores | /day |
| $P_{hc}$ | Preying probability of carnivore on herbivore | Probability of event of preying once herbivore come into the searching area (volume) of carnivore | None (ratio) |
| $P_{cc}$ | Preying probability of carnivore on carnivore | Probability of event of preying once carnivore come into the searching area (volume) of carnivore | None (ratio) |
| $d_h$ | Decreasing constant of herbivore | Ratio of decrease of herbivore biomass caused by respiration, metabolism, and excretion per unit time (day), to biomass of herbivore | /day |
| $d_c$ | Decreasing constant of carnivore | Ratio of decrease of carnivore biomass caused by respiration, metabolism, and excretion per unit time (day), to biomass of carnivore | /day |
| $G_h$ | Relative (internal) growth rate of the herbivore (set) continuously eating plant material per day | The value defined as $(\alpha_h e_h n_p - n_h d_h)/n_h$ | /day |
| $I_h(t)$ | Intensity of herbivory | Volume of plant consumed by herbivore per unit volume ecosystem per day in convergent equilibrium | /day |
| **$I_h$** | Intensity of herbivory in stable equilibrium condition | Volume of plant consumed by herbivore per unit volume ecosystem per day in convergent equilibrium | /day |

Table 2. Comparison between the values predicted from the model and the data from empirical observations about biomasses and annual leaf consumption in forest and grassland ecosystems where insects are main herbivores.

| | Estimated S and $G_h$ (/day) | Herbivore biomass (presented as wet mass including gut content) (mg/m$^2$) | Carnivore biomass (wet mass) (mg/m$^2$) | Annuual consumption by herbivory (%) |
|---|---|---|---|---|
| Prediction from the present model (S is estimated using data from following literature. See text for detail of estimation) | | | | |
| 1. Wasp-Caterpillar system in forest (Morimoto 1960) | $S=1.298\times10^7$ $G_h=0.296$ | 22.9 | 90.6 | 1.14 |
| 2. Carnivorous Bug – Herbivorous Beetle system in forest (Wiedenmann and O'Neil 1991) | $S=3.68\times10^6$ $G_h=0.296$ | 80.6 | 319 | 4.03 |
| 3. Bird-Caterpillar system in tit system (Royama 1970) | $S=7.5\times10^6$ $G_h=0.296$ | 119.6 | 156.8 | 5.98 |
| Data from empirical observations | | | | |
| 1. Biomass of arthropods observed in a forest in North Carolina, USA (Schowalter et al. 1981) | | 120-360 | 40-240 | |
| 2. Biomass of Carnivorous wasps introduced in New Zealand (Thomas et al. 1990) | | | 100-376 | |
| 3. Biomass of insectivorous birds in temperate and tropical forests in North and South America (Ternorgh et al. 1990; Holmes and Sherry 2001) | | | 20-30 | |
| 4. Biomass of herbivorus insects on *Solidago* stands in North America (Meyer et al. 2005) | | 31.2-252 | | |
| 5. Annual consumption rate of plant production in forest ecosytems around the world (Cebrian 1999) | | | | 1.5 (0-14, quart 0-6) |
| 6. Annual consumption rate of plant production in grassland ecosystems around the world (Cebrian 1999) | | | | 23 (0-65, quart 1-43) |
| 7. Annual herbivory in temperate forests (Coley and Barone 1996) | | | | 7.1 |
| 8. Annual herbivory in tropical forests (shade-tolerant species) (Coley and Barone 1996) | | | | 11.1 |
| 9. Area of leaf damage in Permian forests from fossil records (Smith and Nufio 2004; Labandeira and Allen 2007) | | | | 0.25-3.3 |
| 10. Area of leaf damage in Eocene forests from fossil records (Smith and Nufio 2004) | | | | 1.4-2.5 |

Table 3. Comparison between the values predicted from the model and the data from empirical observations about biomasses and annual leaf consumption in African Savannah where large mammals are main herbivores and carnivores.

| | Estimated S and $G_h$ (/day) | Biomass of large herbivorous mammals (presented as wet mass including gut content) (mg/m$^2$) | Population density of wildebeests and zebras (individuals/km$^2$) | Biomass of large carnivorous mammals (wet mass) (mg/m$^2$) | Population density of lions (individuals/km$^2$) | Annual consumption by herbivory (%) |
|---|---|---|---|---|---|---|
| Prediction from the present model (see text for detail of estimations) | | | | | | |
| Lion-Wildebeest/Zebra system in Savanna | S=3.0×10$^4$ $G_h$=0.0002 | 7142 | 35.7 | 26.6 | 0.133 | 65.2 |
| Data from empirical observations | | | | | | |
| Lion-Wildebeest/Zebra system in Serengeti, Masai Mara, and other African Savannah (Coe et al. 1976; McNaughton 1985; Wildt et al. 1987; Packer et al. 2005) | | 8352-30481 | 101.6-121.9 | 24.0-33.8 | 0.129-0.169 | 66(15-95) |

# Figure 1

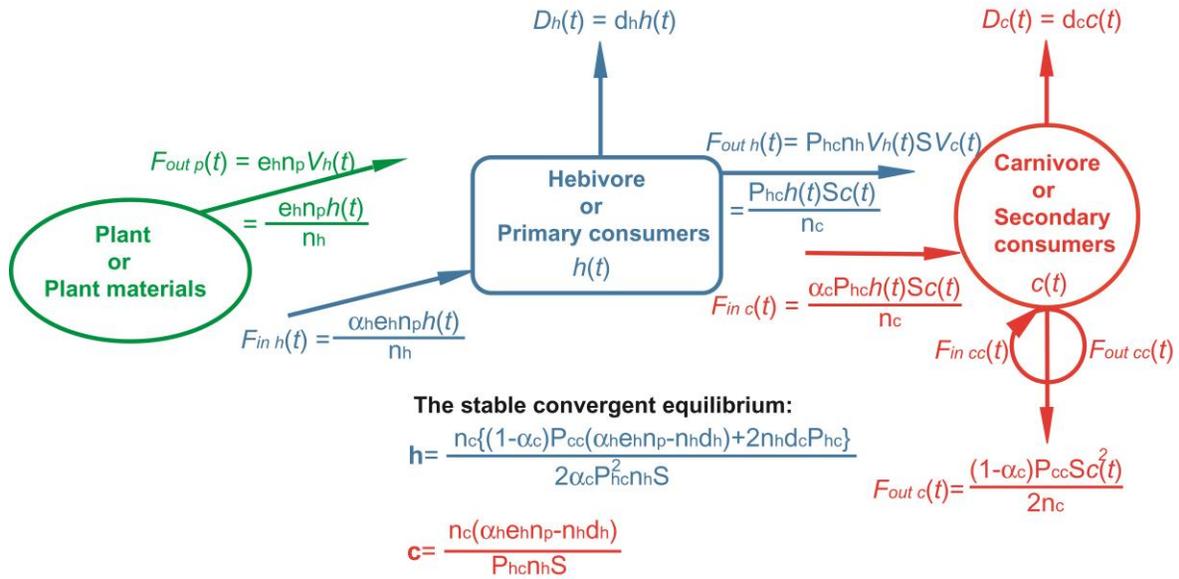

FIG. 1. The structure of the food web model and the convergent stable equilibrium for the biomass of herbivores **h** and carnivores **c**. The diagram indicates the structure of the food web model. The arrows in the diagram show the flow of nutrients (kg protein in this case) per unit volume of ecosystem (m$^3$) per unit time (day), which has a unit of (kg protein/m$^3$day). Green, blue, and red arrows indicate influx and efflux of nutrients (protein) from plants, herbivores, and carnivores, respectively. Influx and efflux balances for both biomass of herbivore $h(t)$ and biomass of carnivore $c(t)$ when equilibrium is reached, and according to this principle, the convergent stable equilibrium of biomass of herbivore **h** and biomass of carnivore **c** is obtained as described below the diagram.

# Figure 2

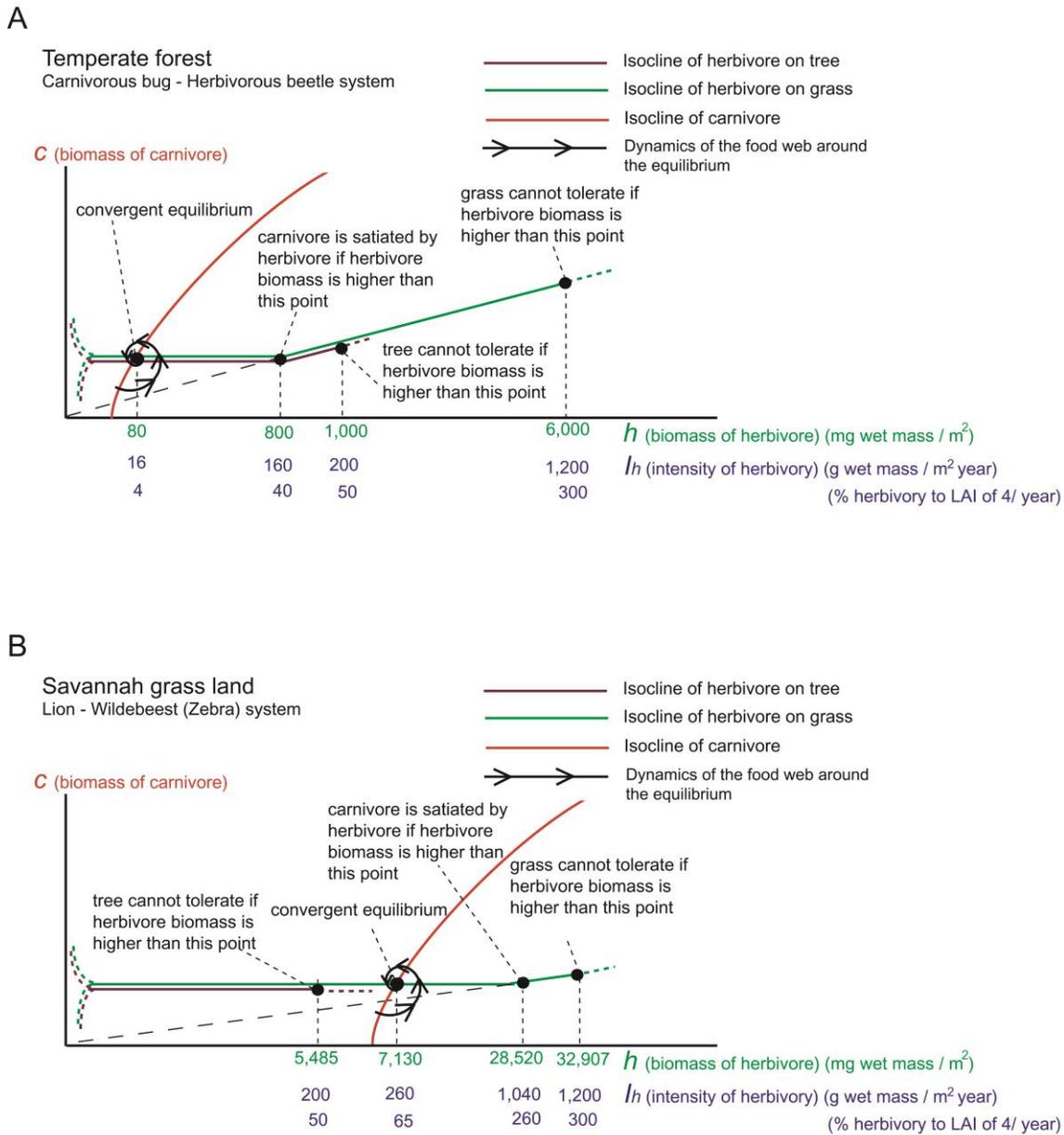

FIG. 2. Graphical representation of food web structures of temperate forest and savannah grassland ecosystems based on isoclines of herbivores and carnivores with predicted values of parameters obtained from the present parameterized mathematical food-web model. (A) A graph showing the situation (parameterized realities) in the temperate forest with carnivorous bug as carnivores and herbivorous beetles as herbivores. Note that $I_h$ (intensity of herbivory) in equilibrium condition, 4 (% herbivory to LAI of 4/year), is smaller than $I_h$ under the condition at which carnivores begin to be satiated by herbivores (40%), and the isocline line for herbivores at the equilibrium point is stable (flat), which suggests that no destabilizing effect that comes from handling time (or satiation of digestive ability in carnivore caused by herbivore) exists. Moreover, at equilibrium, the carnivore isocline inclines to the right because of the existence of intraguild predation of carnivores, which stabilizes the food web system. Therefore, in this case, the graphical representation suggests that the equilibrium is convergent. The $I_h$ (intensity of herbivory) under the equilibrium condition, 4 (% herbivory to LAI of 4/year), is also smaller than both $I_h$ under the condition at which trees can no longer tolerate herbivory and start to decrease in number (50%), and $I_h$ under the condition at which even grass cannot tolerate herbivory anymore and begins to decrease (80%), which means that both forest and grassland can stay green at the equilibrium. Taken together, the details of this order indicate that a stable green forest will be realized (see the text for the detailed rationale). (Grassland may also be realized, but because forest is more competitive in competition for light, forest may finally prevail). (B) A graph showing the situation (parameterized realities) in Savannah grassland with lions as carnivores and wildebeests (or zebras) as herbivores. $I_h$ under equilibrium condition, is smaller than both $I_h$ at the limit condition that carnivores begin to be satiated by herbivores and $I_h$ at limit condition that grass can tolerate, but is larger than $I_h$ at the limit condition that tree can tolerate. This order means that stable forest ecosystem is vulnerable or cannot exist because of the high intensity of herbivory, but that a stable (convergent) green grassland can exist and will be realized. The calculated parameters thus predict the realization of green grassland in this ecosystem (see the text for details). In both case (A) and (B), the $I_h$ at equilibrium is smaller than the $I_h$ at the point of satiation of carnivores by herbivores, and this means that at the equilibrium, the plant isocline is constant (flat) without the existence of satiation that imposes a handling time onto carnivores, which functions as a destabilizing factor. Meanwhile, intraguild predation of carnivores that functions as a stabilizing factor is present in most realistic ecosystems, which is shown in the graph as a rightward inclination of the carnivore isocline. In the absence of a destabilizing factor and the presence of a stabilizing factor, the equilibrium is stable and convergent.